\documentclass{article}
\usepackage[margin=1in,paperheight=11in,paperwidth=8.5in]{geometry}
\usepackage{authblk}
\usepackage{babel}
\usepackage{mathtools}
\usepackage{amssymb} 
\usepackage{float}
\usepackage{enumitem}
\usepackage{amsthm}
\usepackage{amsmath}
\usepackage{amsfonts}
\usepackage{array}
\usepackage{amsopn}
\usepackage{caption}

\usepackage[dvipsnames]{xcolor}

\usepackage{threeparttable}
\usepackage[numbers]{natbib}

\usepackage{enumitem}
\DeclarePairedDelimiterX\Basics[1](){ #1}
\usepackage[utf8]{inputenc}
\usepackage{color}
\usepackage{subcaption}

\usepackage{enumitem}
\usepackage{float}
\usepackage{pifont}
\usepackage{algorithm, algpseudocode, algcompatible}
\usepackage[T1]{fontenc}
\usepackage{etoolbox}
\usepackage{float}
\usepackage{arydshln}
\usepackage{lscape}

\usepackage{multirow}
\usepackage{lineno}
\usepackage{diagbox}
\usepackage{array, booktabs, multirow}
\usepackage{graphicx}

\usepackage{multicol} 
\usepackage{graphicx}
\usepackage{bm}

\usepackage{setspace}
\usepackage{makecell}

\newtheoremstyle{example}{}{}{}{}{\bfseries}{\smallskip}{\newline}{}
\theoremstyle{example}

\newcounter{noteXPctr} 

\usepackage[colorlinks=true,linkcolor=red,filecolor=green,citecolor=blue]{hyperref}

\title{\textbf{Network Momentum across Asset Classes}}

\author[]{Xingyue (Stacy) Pu \thanks{Correspondence to: Xingyue (Stacy) Pu <xingyue.pu@eng.ox.ac.uk>. The authors greatly appreciate the comments and suggestions from Bryan Lim, Kieran Wood and Chao Zhang.}}  
\author[]{Stephen Roberts}
\author[]{Xiaowen Dong}
\author[]{Stefan Zohren}
 
\affil[]{\small Oxford-Man Institute of Quantitative Finance, University of Oxford, Oxford, UK}
\affil[]{\small Department of Engineering Science, University of Oxford, Oxford, UK}

\date{This version: Aug 2023}

\begin{document}

\maketitle

\begin{abstract}

We investigate the concept of network momentum, a novel trading signal derived from momentum spillover across assets. Initially observed within the confines of pairwise economic and fundamental ties, such as the stock-bond connection of the same company and stocks linked through supply-demand chains, momentum spillover implies a propagation of momentum risk premium from one asset to another. The similarity of momentum risk premium, exemplified by co-movement patterns, has been spotted across multiple asset classes including commodities, equities, bonds and currencies. However, studying the network effect of momentum spillover across these classes has been challenging due to a lack of readily available common characteristics or economic ties beyond the company level. In this paper, we explore the interconnections of momentum features across a diverse range of 64 continuous future contracts spanning these four classes. We utilise a linear and interpretable graph learning model with minimal assumptions to reveal the intricacies of the momentum spillover network. By leveraging the learned networks, we construct a network momentum strategy that exhibits a Sharpe ratio of 1.5 and an annual return of 22\%, after volatility scaling, from 2000 to 2022. This paper pioneers the examination of momentum spillover across multiple asset classes using only pricing data, presents a multi-asset investment strategy based on network momentum, and underscores the effectiveness of this strategy through robust empirical analysis.
 
\end{abstract}

\bigskip



\newpage
\tableofcontents
\newpage

\section{Introduction}

\subsection{From Momentum Spillover to Network Momentum}
Momentum as a risk premium in finance refers to the persistent abnormal returns demonstrated by the propensity of winning assets to continue winning and losing assets to continue losing. It has been discovered and investigated across a wide range of asset classes, prediction horizons and time periods \cite{asnessValueMomentumEverywhere2013,bazDissectingInvestmentStrategies2015, hurstCenturyEvidenceTrendFollowing2017}. The momentum premium, interestingly, is not confined to individual assets. There are instances where the performance of one asset appears to lead a similar performance of another, suggesting a propagation of momentum across assets. Such a phenomenon was first documented as \textit{momentum spillover} by \citet{gebhardtStockBondMarket2005}.  \par


The concept of momentum spillover was initially explored in the context of pairwise similarity between two assets, such as those between the bond and stock of the same firm. By examining the lead-lag relationship, \citet{gebhardtStockBondMarket2005} discovered that firms with high equity returns in the past tend to earn high bond returns in the future. A similar spillover effect from equity to high-yield corporate bonds was further confirmed by \citet{haesenMomentumSpilloverStocks2017a}. It is important to note that while many studies on momentum spillover investigate the lead-lag effect in asset returns, these two terms should not be conflated. Momentum spillover is not simply about any stock returns predicting bond returns, but specifically about stock returns that exhibit a momentum premium \cite{gebhardtStockBondMarket2005}. Specifically, it essentially denotes when an asset has previously shown a good performance, its connected assets will perform well, suggesting a long position. Conversely, if an asset has displayed a weak performance in the past, its connected assets are expected to underperform, suggesting a short position.  \par

Beyond the bond-stock connection through the same underlying firm, many studies have identified momentum spillover in the equity returns of firms linked by economic and fundamental ties. A prime example is the supply-demand chain, where firms closely linked along the chain often experience similar cash flow shocks \cite{cohenEconomicLinksPredictable2008b, menzlyMarketSegmentationCrosspredictability2010}. This similarity, stemming from direct trading relationships or shared market influences, contributes to cross-predictability in returns. The phenomenon of momentum spillover is not limited to a single firm or industry. It permeates across firms and industries that share similar fundamentals. This becomes particularly evident when we consider contexts that focus on the similarity of firms. For instance, momentum spillover has been observed among firms that belong to the same industry \cite{moskowitzIndustriesExplainMomentum1999,grobysRiskmanagedIndustryMomentum2018}, operate in the same market segments \cite{cohenComplicatedFirms2012}, or are located in the same region \cite{parsonsGeographicLeadLagEffects2020}. In the realm of technological innovation, firms sharing the similar patents exhibit a lead-lag effect, where the returns of technology-linked firms strongly predict the returns of the target firms \cite{leeTechnologicalLinksPredictable2019}. Moreover, alternative news data and online search data has been used to identify firm similarities \cite{leeSearchPeerFirms2016}, and sell-side analyst coverage can serve as a strong and versatile proxy for fundamental linkages between firms \cite{aliSharedAnalystCoverage2020}, thereby capturing the unified phenomenon of momentum spillover. These examples underscore the pervasive nature of momentum spillover across diverse economic and fundamental linkages. \par

Building upon the concept of momentum spillover in pairwise relationships, further studies have broadened the scope to consider more complex network effects within a universe of stocks. In these networks, firms are not just economically and fundamentally linked in node pairs, but are part of a larger, interconnected system. This network perspective allows for a more comprehensive understanding of how momentum can propagate through a system of interconnected assets, beyond just pairwise connections. For example, \citet{yamamotoMomentumInformationPropagation2021} modified the customer momentum strategy in \cite{cohenComplicatedFirms2012} by applying network theory, specifically using edge betweenness centrality to calculate the importance of supplier-customer relationships. \par

Based on the understanding of the interconnections between assets and momentum spillover, we can derive a unique momentum signal for a target asset. This is achieved by averaging the time-series momentum characteristics of its connected assets, with the weights determined by the strength of the connections. We refer to this type of momentum signal as \textit{Network Momentum}. For instance, in the study by \citet{aliSharedAnalystCoverage2020}, stocks were sorted into quintiles based on the average past one-month return of their connected firms in the analyst coverage network. The portfolio that longs the top quintile and shorts the bottom quintile exhibits a strong monotonic relationship between past returns of connected firms and future returns of the stocks, thereby demonstrating the significant alpha generation potential of network momentum. Oppositely, we adopt a term, \textit{Individual Momentum} \cite{moskowitzIndustriesExplainMomentum1999}, to refer to those momentum characteristics of an asset that are constructed from its own past returns. It is worth noting that cross-sectional momentum \cite{jegadeeshReturnsBuyingWinners1993,rouwenhorstInternationalMomentumStrategies1998,pohBuildingCrossSectionalSystematic2021,tanSpatioTemporalMomentumJointly2023}, as a result of ranking the individual momentum of a universe of assets, are not considered as network momentum. As the aforementioned example \cite{aliSharedAnalystCoverage2020}, it is also possible to construct a cross-sectional momentum portfolio by ranking the network momentum. \par
 
\subsection{Network Momentum across Asset Classes}
Although momentum premium has been confirmed in multiple asset classes, such as commodities, bond and currencies \cite{asnessValueMomentumEverywhere2013}, the concept of network momentum across asset classes, especially its cross-class predictability, has not been exhaustively investigated to the best of our knowledge. Preliminary evidence indeed suggests the existence of co-movement patterns in the momentum of diverse asset classes \cite{asnessValueMomentumEverywhere2013}. Notable studies have identified momentum spillover between bonds and equities \cite{gebhardtStockBondMarket2005, haesenMomentumSpilloverStocks2017a, pitkajarviCrossassetSignalsTime2020, declerckTrendFollowingSpilloverEffects2019}, and the predictive power of historical changes in crude oil index volatility on global stock markets \cite{fernandez-perezCrossassetTimeseriesMomentum2022}. Moreover, significant time-series spillover effects from equities to FX have been observed \cite{declerckTrendFollowingSpilloverEffects2019}, and the impact of currency news announcements on bond returns in emerging markets has also been confirmed \cite{yamaniCurrencyNewsInternational2021}. Yet, these studies do not fully capture the complexity of the interconnected system across asset classes, as they primarily focus on pairwise relationships. This approach overlooks the potential for momentum spillover within a larger, more complex network of assets. 
Despite the spillover being well examined by sophisticated statistical tests, the development and practical application of network momentum signals, particularly in portfolio construction, have not been fully realised. \par

However, building such a network and establishing network momentum across asset classes presents unique challenges due to the complexity of interconnections and the limited availability of readily accessible information. This challenge is particularly evident in commodities, where the absence of firm-like economic and fundamental linkages adds complexity to establishing connections \cite{xuCommodityNetworkPredictable, hanLeadlagRelationsCommodity2022}. For instance, commodities such as cotton may not share direct economic or customer-based relationships with other assets like oil. Additionally, different asset classes possess distinct features and characteristics, further contributing to the complexity of identifying commonalities to draw linkage. This inherent heterogeneity necessitates the use of innovative tools and novel research approaches to apply network momentum at a cross-class level. \par

To address these challenges, we propose a novel approach using a graph learning model \cite{kalofoliasHowLearnGraph2016} to infer dynamic networks among 64 continuous futures contracts across four asset classes: commodities, equities, fixed income (FI), and foreign currencies (FX). From a graph signal processing perspective \cite{dongLearningGraphsData2019, dongLearningLaplacianMatrix2016, kalofoliasHowLearnGraph2016}, graph learning involves estimating a graph adjacency or Laplacian matrix with a model assumption that graph signals mainly consists of low frequency components in the graph spectral domain, i.e. low-pass graph signals \cite{ramakrishnaUserGuideLowPass2020}. It is therefore expected that graph signals have a slow variation over the resulting graph. The variation is measured by the Laplacian quadratic term \cite{dongLearningGraphsData2019, mateosConnectingDotsIdentifying2019}. In our proposal, each asset is a node, the interconnections of assets are represented as a graph adjacency matrix, and graph signals are a collection of individual momentum characteristics of every asset. The graph signals are derived directly from asset pricing data, bypassing the absence of readily economic and fundamental ties beyond the company level. The graph learning model \cite{kalofoliasHowLearnGraph2016} solves a convex optimisation problem to infer a graph adjacency matrix, with the primary objective to minimise the aforementioned Laplacian quadratic term. This approach, while exploring linear relationships between node observations, effectively considers the interconnected system as a whole by leveraging the low-pass properties of graph signals in the spectral domain \cite{ramakrishnaUserGuideLowPass2020}. 
The learned graph adjacency matrix essentially acts as a network\footnote{In this paper, we use \textit{graph} and \textit{network} interchangeably. A \textit{node} is referred as to an \textit{asset}.} with inherent properties that allow for the construction of network momentum, as illustrated in previous examples. Specifically, it maintains only non-negative edge values, reflecting the strength of similarity in momentum features between paired assets, and no self-connections. The resulting graph structure provides an interpretable representation of complex relationships across assets and asset classes, capturing momentum spillover patterns that are directly learned from historical momentum observations. This marks a difference from correlation-based or regression-based statistical models in the literature, which typically rely on the lead-lag effect to study momentum spillover \cite{gebhardtStockBondMarket2005, xuCommodityNetworkPredictable}.  \par


Once the networks are established from graph learning, we propose a linear regression model to devise the network momentum strategy. For each asset, the covariates are network momentum features, which are weighted average of its connected assets' individual momentum features, with edge values as weights. The model aims to predict an asset's future trend, targeted by its future 1-day volatility-scaled return. We train a single model across all assets. We use eight momentum features as per the literature \cite{limEnhancingTimeSeries2020}, including volatility-scaled returns and normalised moving average crossover divergence (MACD) over different time spans. This approach is anticipated to offer improvements over the previously mentioned model-free network momentum strategies \cite{aliSharedAnalystCoverage2020, yamamotoMomentumInformationPropagation2021}, which use the average past returns of connected assets as the predictive return of a target asset. Notably, these strategies might not fully account for risk characteristics, such as the volatilities or skewness of the predictive return distribution, potentially increasing exposure to large downside moves \cite{danielMomentumCrashes2016,barrosoMomentumHasIts2015}. 

Additionally, regression on diverse momentum features might shed some light on network momentum reversals. The phenomenon of short-term reversal has been well-established in individual momentum \cite{jegadeeshEvidencePredictableBehavior1990a, daCloserLookShortTerm2014, kellyUnderstandingMomentumReversal2021}. This phenomenon suggests that assets that have performed well in the recent past (e.g., over the last month) tend to underperform in the near future (e.g., over the subsequent month), and vice versa. However, there lacks a unified and comprehensive understanding of the persistence of such reversals within the realm of network momentum. On the one hand, certain research has identified significant alpha from a signal that combines the short-term reversal with network momentum \cite{israelsenDoesCommonAnalyst2016, shahrurReturnPredictabilitySupply2010}. On the other hand, studies \cite{moskowitzIndustriesExplainMomentum1999, aliSharedAnalystCoverage2020, parsonsGeographicLeadLagEffects2020} find little to no evidence of reversals in network momentum, contrasting sharply with individual momentum \cite{jegadeeshEvidencePredictableBehavior1990a}. Failing to account for reversals, should they exist, can compromise portfolio performance. In this paper, we propose an alternative solution that apply linear regressions on those momentum features across different time spans. This approach not only addresses the potential pitfalls of overlooking reversals but also provides valuable insights into the nature of network momentum and potential reversals, through a detailed examination of regression coefficient significance and signs. \par
  

We conducted a backtest of the proposed network momentum strategy on 64 continuous futures contracts spanning four asset classes: Commodities, Equities, Fixed Income, and Currencies, over an out-of-sample period from 2000 to 2022. The strategy demonstrated impressive profitability, achieving an annual return of 22\% and a Sharpe ratio of 1.51, after volatility scaling. Moreover, the strategy effectively managed risk, exhibiting lower downside deviation and durations. Remarkably, it showed low correlation with individual momentum \cite{limEnhancingTimeSeries2020}, suggesting that the incorporation of networks can generate novel signals. \par

To ensure the strategy's robustness, we conducted various examinations, including diversification and turnover analysis, and also a deep investigation into the topological structure and temporal stability of the learned graphs. We also evaluated the impact of inter- and intra-class connections. Our investigation underscores the critical role of inter-class connections in shaping network momentum signals. These connections bridge various asset classes, integrating their distinct momentum features into a cohesive strategy and thus enhancing predictive power. Notably, the advantages of these connections extend beyond cross-asset class trading. Even in single-asset trading, inter-class connections can offer benefits due to the propagation of momentum effects, emphasising the value of comprehending network structure and dynamics in optimising trading strategies. \par

\subsection{Main Contributions}

The contributions of this paper are threefold. Firstly, it pioneers the examination of momentum spillover across multiple asset classes using graph learning and pricing data only, while introducing the concept of network momentum, a novel trading signal. Secondly, it presents a methodology for portfolio construction across different asset classes based on these signals, and proposes a linear regression-based trading strategy, thereby enhancing the existing model-free strategies by considering the risk characteristics and reversals. Lastly, through extensive empirical testing and graph topological analysis, this paper underscores the effectiveness of the proposed strategy, highlights the pivotal role of cross-asset and cross-class connections in forming network momentum signals, and offers insights into the topological structure and temporal dynamics of these interconnections. This paper thus delivers considerable value in cross-asset trading strategy development, especially across various asset classes. \par

\section{Data}

\subsection{Dataset}

Our raw dataset contains the daily prices of 64 highly liquid ratio-adjusted continuous futures contracts. These contracts are extracted from the Pinnacle Data Crop CLC Database\footnote{Database obtained from \href{https://pinnacledata2.com/clc.html}{https://pinnacledata2.com/clc.html}}. The contracts represent four classes of assets: Commodities, Equities, Fixed Income (FI), and Currencies (FX). The daily price data are available from 1990 to 2022, although different assets may have different spans in terms of available prices. In Appendix, Table \ref{table:universe} provides a list of all the assets used in the backtest. \par


\subsection{Momentum Features}
\label{sec:data-features}
The following eight individual momentum features \cite{limEnhancingTimeSeries2020} are calculated from the raw pricing data as the input of the graph learning model to infer the network between 64 assets, and also the individual momentum features as the information to propagate to obtain network momentum. 
\begin{itemize}
    \setlength\itemsep{-0.5em}
    \item \textbf{Volatility-scaled returns} over the past 1-day, 1-month, 3-month, 6-month and 1-year periods, denoted as $\frac{r_{t-\Delta:t}}{\sigma_t \sqrt{\Delta}}$ for asset $i$, where $\Delta = \{1, 21, 63, 126, 252\}$ days. 
    The daily volatility $\sigma_t$ is estimated using an exponential weighted moving standard deviation (EWMstd) with a 60-day span on daily returns. This method gives more weight to recent observations, using a smoothing factor $\alpha = \frac{2}{60+1}$. The 60-day span refers to the window over which the weights decay, making this a responsive measure of market risk. The definition of EWMstd can be found in Appendix \ref{sec:appendix-math-preprocs}.
    
    \item \textbf{Normalised MACD} indicators $y_{i,t}(S,L)$ for asset $i$ at day $t$, developed by \citet{bazDissectingInvestmentStrategies2015}, with three different combinations of short and long time scales $(S_k, L_k) \in \{(8,24), (16, 48), (32, 96)\}$, such that
    \vspace{-1em}
    \begin{align}
    & \text{MACD}(i, t, S, L) = m(i,t,S) - m(i, t, L)  \label{eq:macd_ewm}\\
    & \text{MACD}_{\text{norm}}(i, t, S, L) =  \frac{\text{MACD}(i, t, S, L)}{\text{std}(p_{i, t-63:t})} \\
    & y_{i,t}(S,L) =  \frac{\text{MACD}_{\text{norm}}(i, t, S, L) }{\text{std}(\text{MACD}_{\text{norm}}(i, t-252, S, L) )} \label{eq:macd}
    \end{align}%
    where $m(i,t,J)$ is the exponentially weighted moving average of the prices of asset $i$ at time $t$, with a scale $J$ such that the smoothing factor $\alpha = 1/J$ (see definition in Appendix \ref{sec:appendix-math-preprocs}). Here $\text{std}(p_{i, t-63:t})$ is the 63-day rolling standard deviation of the prices of asset $i$. Similarly, the denominator of Eq.\eqref{eq:macd} is the 252-day rolling standard deviation of the normalised MACD.
\end{itemize}
Aggregating the above, we have a feature matrix $\mathbf{U}_t = [\mathbf{u}_{1,t}, \dots, \mathbf{u}_{i,t}, \dots, \mathbf{u}_{N_t, t}]^T \in \mathbb{R}^{N_t \times 8}$ for every trading day $t$ in the backtest period,  where $N_t$ is the number of assets. We adopted a data winsorisation strategy for momentum features in order to mitigate the influence of outliers. Specifically, each feature of each asset was capped and floored to fall within a range defined by five times its exponentially weighted moving standard deviations from its corresponding exponentially weighted moving average. This computation was carried out using a half-life of 252 days. \par

\section{Network Momentum}

\subsection{Graph Learning} \label{sec:graph_learning}

The initial step of constructing network momentum necessitates the development of networks to represent the interconnection between assets. The network is expected to reflect the historical similarities in individual momentum between assets, which form a cornerstone for the meaningful propagation of individual momentum from one asset to its connected assets. Importantly, networks should be dynamic, flexibly adjusting to the evolving financial market regimes and authentically capturing the ever-changing interconnections. Moreover, these networks should have key characteristics: non-negative edge weights -- to reflect the relative importance of one asset to another; sparsity -- to ensure an asset is linked only to certain essential assets, thereby safeguarding the momentum propagation from potential noise contamination. Moreover, the model should be interpretable, providing crucial understanding in terms of financial data. Graph learning \cite{kalofoliasHowLearnGraph2016, dongLearningLaplacianMatrix2016, mateosConnectingDotsIdentifying2019} is applied to achieve this goal.  \par

On each trading day $t$, we stack the momentum feature matrices, a.k.a $\mathbf{U}_t$ defined in Section \ref{sec:data-features}, over a lookback window of $\delta$ days. This produces a matrix $\mathbf{V}_t \in \mathbb{R}^{N_t \times 8\delta}$, where each row corresponds to assets consistently available throughout the lookback window. The columns represent a concatenation of 8 momentum features over $\delta$ days. To estimate the graph adjacency matrix $\mathbf{A}_t \in \mathbb{R}^{N_t \times N_t}$, which signifies the sought-after network among assets, the graph learning model \cite{kalofoliasHowLearnGraph2016} solves the following convex optimisation problem:
\begin{equation}
\label{eq: gl}
\begin{aligned}
    \text{(graph learning)} \quad \min_{\mathbf{A}_t} & ~~  \text{tr}\Big(\mathbf{V}_t^{\top}(\mathbf{D}_t -\mathbf{A}_t) \mathbf{V}_t \Big)- \alpha \mathbf{1}^{\top} \log (\mathbf{A}_t \mathbf{1})+ \beta ||\mathbf{A}_t||_F^2  \\
    s.t. & ~~  \mathbf{A}_{ij, t} = \mathbf{A}_{ji, t}, ~~ \mathbf{A}_{ij, t} \geq 0 ~~ \forall i \neq j
\end{aligned}
\end{equation}%
where $\mathbf{D}_t$ is a diagonal matrix with $\mathbf{D}_{ii,t} = \sum_j \mathbf{A}_{ij,t}$. The adjacency matrix $\mathbf{A}_t$ represents the network at day $t$ for constructing network momentum, with the $ij$-th entry $\mathbf{A}_{ij, t}$ measuring the strength of similarity of individual momentum between asset $i$ and asset $j$. In the objective function, the first trace
term measures the spectral variations of $\mathbf{V}_t$ on the learned graph adjacency matrix $\mathbf{A}_t$, encouraging connections between nodes with similar features. It is derived from Laplacian smoothness under the mild assumption that each column of $\mathbf{V}_t$ is a low-pass graph signal \cite{ramakrishnaUserGuideLowPass2020}. Additionally, this term suggests that the model is linear in the dot product of the $8\delta$-dimensional momentum feature series of two assets. The log and $\ell_2$ terms act as topological regularisation in the optimisation problem, enforcing graph connectivity to prevent isolated nodes and ensure a smooth edge weight distribution. The constraints guarantee
the learned graph adjacency matrix is symmetric and non-negative, which aligns with our expectations of a network for constructing network momentum as aforementioned. \par

There are two hyperparameters $\alpha$ and $\beta$. Together they control the topological properties of learned graphs, such as sparsity. The smaller the values of $\alpha$ and $\beta$, the sparser the resulting graph will be, and hence every target asset will receive information form just a few other assets. In the propagation of individual momentum characteristics, too many connections will introduce noise, while too few might not capture the essential connections. Therefore, the choice of $\alpha$ and $\beta$ can largely affect the performance of network momentum strategies. We adopt a discrete grid search on in-sample data to determine the values of them. \par
 
In our empirical analysis, we combine $K=5$ distinct graphs learned from $\mathbf{V}_t$ from five different lookback windows such that $\delta \in \{252, 504, 756, 1008, 1260\}$ trading days as follows:
\begin{equation} \label{eq: graph_ensemble}
    \text{(graph ensemble)} \quad \bar{\mathbf{A}}_t = \frac{1}{K}\sum_{k=1}^{K} \mathbf{A}^{(k)}_t.
\end{equation}%
The construction of graphs with annual information aligns with the literature on network momentum \cite{aliSharedAnalystCoverage2020,cohenEconomicLinksPredictable2008b}. The purpose of this ensemble is to reduce the variance of the learned edge weights, which could be unstable due to potential changes in the market regime. 
Our empirical results also indicate that this ensemble strategy helps improve portfolio performance in terms of profitability and reduces turnover (see Section \ref{sec:robust_lookback}). \par

To mitigate the effects of scale differences in constructing network momentum, which may arise due to the difference in the number of connections certain assets have -- with some connected to numerous other assets and others only to a few -- we also apply a graph normalisation as follows:
\begin{equation}
    \text{(graph normalisation)} \quad \Tilde{\mathbf{A}}_t = \bar{\mathbf{D}}_t^{-1/2}\bar{\mathbf{A}}_t \bar{\mathbf{D}}_t^{-1/2}
\end{equation}%
where $\bar{\mathbf{D}}_t$ is a diagonal matrix with $\bar{\mathbf{D}}_{ii,t} = \sum_j \bar{\mathbf{A}}_{ij,t}$.

\subsection{Portfolio Construction}

With the learned graphs from the previous section, we can construct the network momentum features for each asset by propagating the individual momentum of its connected assets as follows:
\begin{equation}
\label{eq: network_momentum}
      \text{(network momentum features)} \quad \Tilde{\mathbf{u}}_{i,t} = \sum_{j \in \mathcal{N}_t(i)} \Tilde{\mathbf{A}}_{ij, t}  \mathbf{u}_{j,t}  
\end{equation}%
where $\mathcal{N}_t(i)$ is the set of assets connected to asset $i$ such that $\mathbf{A}_{ij, t} \neq 0$, and $\mathbf{u}_{j,t}$ is the vector of eight individual momentum features in Section \ref{sec:data-features} for asset $j$ at day $t$. The propagation mechanism simply takes a weighted average of individual momentum features, the weights being the strength of the edges, which aligns with most literature \cite{cohenEconomicLinksPredictable2008b, aliSharedAnalystCoverage2020, yamamotoMomentumInformationPropagation2021}. Our method differs in that we utilise eight risk-adjusted individual momentum features, rather than a single and vanilla momentum feature, which is often represented by past-month returns \cite{aliSharedAnalystCoverage2020, yamamotoMomentumInformationPropagation2021}. Next, we estimate the momentum trend with an OLS linear regression model such that
\begin{equation}
     \text{(network momentum)} \quad y_{i,t} = \Tilde{\mathbf{u}}_{i,t}^{T} \boldsymbol{\beta} + b
\end{equation}%
where $\boldsymbol{\beta}$ is an eight-dimensional vector of coefficients that corresponds each of the network momentum features and $b$ is the intercept term. These coefficients are estimated cross-sectionally, i.e. with in-sample data of all assets $\{\mathcal{D}_t\}_{t=1}^T$ of a length of $T$ days, and $\mathcal{D}_t = \{(\tilde{\mathbf{u}}_{i,t}, y_{i,t})\}_{i = 1}^{N_t}$. The prediction target of $y_{i,t}$ is the future 1-day volatility-scaled return, ${r_{i, t:t+1}}/{\sigma_{i,t}}$, as defined in Section \ref{sec:data-features}. \par

We employ linear regression instead of the vanilla individual momentum obtained from past-month returns, because the latter does not account risk characteristics, potentially exposing the positions to large downside moves. Another consideration is the reversal effect; while momentum reversal is well-documented in individual momentum strategies, the existence of reversal in network momentum has not been universally confirmed. The regression coefficients can shed some light on this by examining the sign and significance of these momentum features over different time spans, which can also potentially enhance performance. This approach aligns with recent literature on machine learning momentum \cite{limEnhancingTimeSeries2020}. \par

The daily return of a long/short portfolio, labelled as \textbf{GMOM}, based on the network momentum is defined as \cite{limEnhancingTimeSeries2020}:
\begin{equation}
\label{eq: TSMOM}
r_{t:t+1}^{\text{portfolio}} := \frac{1}{N_t}\sum_{i=1}^{N_t} x_{i,t} \frac{\sigma_{\text{tgt}}}{\sigma_{i,t}} r_{i, t:t+1}
\end{equation}%
where $x_{i,t}  = \text{sign}(y_{i,t})$ indicate the long/short position, $r_{i, t:t+1}$ is the daily return of asset $i$. With the target annualised volatility $\sigma_{\text{tgt}}$, the asset return is scaled by its annualised realised volatility $\sigma_{i,t}$. Here, $\sigma_{\text{tgt}} = 0.15$ following literature standards \cite{limEnhancingTimeSeries2020}, and $\sigma_{i,t}$ is estimated from an exponential weighted moving standard deviation with a 60-day span on daily returns (see definition in Appendix \ref{sec:appendix-math-preprocs}). \par

\section{Backtest}
\label{sec:4backtest}

\subsection{Setup}

\paragraph{Strategy Candidates}
For comparison against traditional individual momentum strategies, we compare the proposed strategy \textbf{GMOM} against the following reference benchmarks:
\begin{itemize}
    \setlength\itemsep{-0.5em}
    \item{\textbf{Long Only}} takes a consistent long position $x_{i,t} = 1$ in Eq.\eqref{eq: TSMOM} with volatility scaling for all the assets. This also represents the market benchmark. 
    \item{\textbf{MACD}} is a model-free individual momentum strategy, proposed by \citet{bazDissectingInvestmentStrategies2015}, which takes the average of three normalised MACD indicators in Eq.\eqref{eq:macd} of different time scales combinations such that 
    \begin{equation}
         x_{i,t} = \frac{1}{3} \sum_{k=1}^3 \phi\big( y_{i,t}(S_k,L_k)\big)
    \end{equation}%
    where $(S_k, L_k) \in \{(8,24), (16, 48), (32, 96)\}$ and $\phi(y) = \frac{y\exp(-y^2/4)}{0.89}$ is a position scaling function. 
    \item{\textbf{LinReg}} stands for Linear Regression with the eight individual momentum features in Section \ref{sec:data-features} as input such that 
    \begin{equation}
         \text{(individual momentum)} \quad y_{i,t} = \mathbf{u}_{i,t}^{T} \boldsymbol{\beta} + b.
    \end{equation}%
\end{itemize}%

\paragraph{Optimisation Details}

The models were trained every 5 years. This process involved coefficient optimisation and hyperparameter tuning using all preprocessed data up to the point of re-calibration. Afterwards, these optimised models were used to generate trading signals for the subsequent 5 years, using out-of-sample data. To ensure an adequate number of training samples, the first backtest period comprised of 1990-1999 for training, followed by 2000-2004 for testing. Likewise, the second period included 1990-2004 for training, and then 2005-2009 for testing, and so on. Therefore, we have an aggregated 22-year out-of-sample period from 2000 to 2022. Note that the last period include 1990-2019 for training, and then 2020-2022 for testing. For each backtest period, the most recent 10\% of training data (a.k.a. in-sample data) was set aside as a validation set for hyperparameter grid search in graph learning. The discrete grid for $\alpha$ and $\beta$ in Eq.~\eqref{eq: gl} is $\{0.0001, 0.0005, 0.001, 0.005, 0.01, 0.05, 0.1, 0.5, 1, 5, 10\}$. The optimisation problem of graph learning, convex but not differentiable, is solved numerically with MOSEK in the CVXPY interface \cite{diamond2016cvxpy}. The linear regression has a closed-form analytical solution. \par

\subsection{Portfolio Performance} \label{sec:4backtest-portftolio-performance}
\begin{figure}[h] 
    \centering
    \begin{subfigure}[]{0.49\textwidth}
        \includegraphics[width=1\textwidth]{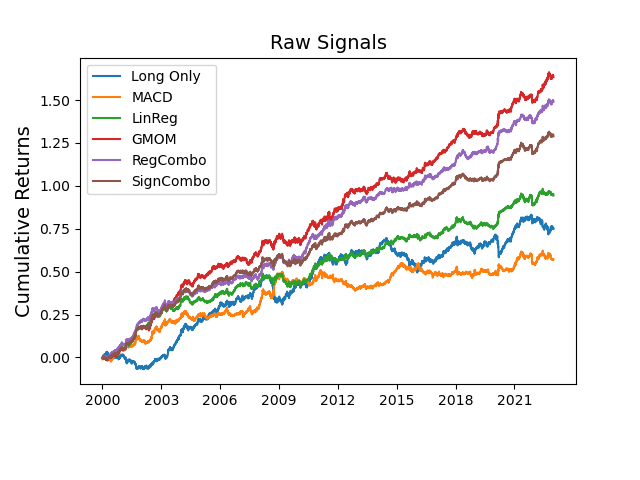}
        \caption{Raw signals}
        \label{subfig:cumulative_returns_raw}
    \end{subfigure}
    \hspace{0.1cm}
    \begin{subfigure}[]{0.49\textwidth}
        \includegraphics[width=1\textwidth]{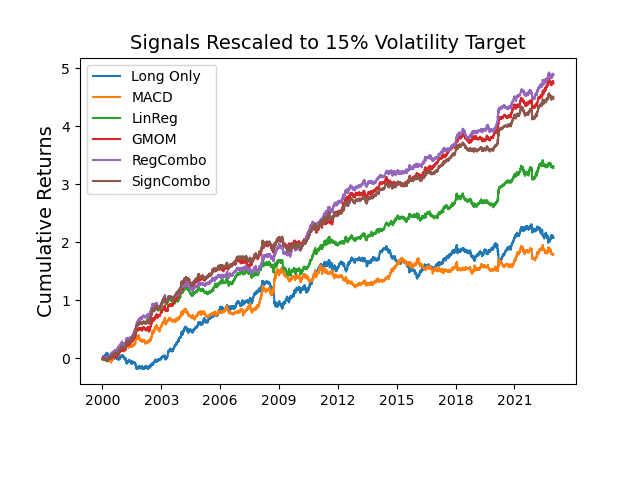}
        \caption{Signals scaled to 15\% volatility target}
        \label{subfig:cumulative_returns_scaled}
    \end{subfigure}
\caption{The cumulative daily returns of the proposed strategy GMOM, three reference strategies (Long Only/MACD/LinReg) and two mixed strategies from LinReg and GMOM (RegCombo/SignCombo - see definition in Section \ref{sec:backtest_diverse}) in different colours for the entire out-of-sample period from 2000 to 2022.}
\label{fig:cumulative_returns}
\end{figure}

In evaluating the portfolio performance, we use the following annualised metrics in Table \ref{tab:bt_perf_table_combined}:
\begin{itemize}
    \setlength\itemsep{-0.5em}
    \item{\textbf{Profitability}}: expected return and hit rate (the percentage of days with positive returns across the test period).
    \item{\textbf{Risk}}: volatility (vol.), downside deviation, maximum drawdown (MDD) and MDD duration (the percentage of days with MDD across the test period).
    \item{\textbf{Overall Performance}}: Sharpe ratio (expected return$/$vol.), Sortino ratio (expected return$/$downside deviation), Calmar ratio (expected return$/$MDD), and the average profits over the average loss ($\frac{\text{Avg. P}}{\text{Avg. L}}$).
\end{itemize}%
In Panel A of Table \ref{tab:bt_perf_table_combined}, we report the performance of portfolio constructed from raw signals. In order to have a better comparison between different strategies, we apply an additional layer of volatility scaling at the portfolio level to an annualised volatility target of 15\% and report the above evaluation metrics in Panel B of Table \ref{tab:bt_perf_table_combined}. The cumulative returns of each strategies are plotted, with raw signals in Figure \ref{subfig:cumulative_returns_raw}, and volatility-scaled signals in Figure \ref{subfig:cumulative_returns_scaled}.

\begin{table*}[t]
\caption{Portfolio Performance Metrics}
\label{tab:bt_perf_table_combined}
\small
\centering
\begin{threeparttable}
\begin{tabular}{lccccccccccc}
\toprule
{} &  return &    vol. &  Sharpe &  \makecell{downside \\ deviation} & \makecell{MDD} &  \makecell{MDD \\ duration} &  Sortino &  Calmar &  hit rate &  $\frac{\text{Avg. P}}{\text{Avg. L}}$ \\
\midrule
\multicolumn{11}{c}{Panel A: Raw Signals} \\
\midrule
Long Only &0.032 & 0.054 &   0.599 &               0.039 &         0.200 &                 15.7\% &    0.840 &   0.157 &    53.6\% &                 0.954 \\
MACD       & 0.025 & 0.048 &   0.516 &               0.035 &         0.121 &                 25.5\% &    0.708 &   0.197 &    53.1\% &                 0.967 \\
LinReg     & 0.041 & \bf{0.043} &   0.943 &               0.031 &         0.080 &                 10.9\% &    1.334 &   0.506 &    53.7\% &                 1.014 \\
GMOM      & \bf{0.070} & 0.052 &   1.363 &               0.035 &         0.071 &                  6.9\% &    2.026 &   \bf{1.013} &    \bf{55.2\%} &                 1.024 \\ [0.1cm]
\multicolumn{11}{l}{Mixed Signals of LinReg \& GMOM:}  \\
$~$ RegCombo  & 0.064 & 0.046 &   \bf{1.403} &               \bf{0.030} &         \bf{0.067}  &                  \bf{5.3\%} &    \bf{2.131} &   0.972 &    54.6\% &                 \bf{1.054} \\
$~$ SignCombo & 0.056 & 0.043 &   1.286 &               0.031 &         0.073 &                  7.6\% &    1.814 &   0.772 &    55.2\% &                 1.014 \\
\midrule
\multicolumn{11}{c}{Panel B: Signals Rescaled to 15\% Target Volatility}  \\
\midrule
Long Only & 0.108 & 0.148 &   0.734 &               0.096 &         0.633 &                 15.7\% &    1.133 &   0.162 &    53.6\% &                 0.972 \\
MACD      & 0.103 & 0.147 &   0.699 &               0.095 &         0.342 &                 21.6\% &    1.076 &   0.281 &    53.1\% &                 0.990 \\
LinReg    & 0.165 & 0.146 &   1.128 &               0.094 &         0.285 &                 10.9\% &    1.764 &   0.586 &    53.7\% &                 1.036 \\
GMOM      & 0.222 & 0.147 &   1.511 &               0.092 &         \bf{0.199} &                  6.9\% &    2.422 &   1.179 &    \bf{55.2\%} &                 1.038 \\ [0.1cm]
\multicolumn{11}{l}{Mixed Signals of LinReg \& GMOM:}  \\
$~$ RegCombo  & \bf{0.226} & 0.147 &   \bf{1.536} &               0.091 &         0.201 &                  \bf{5.3\%} &   \bf{2.470} &   \bf{1.194} &    54.6\% &                 \bf{1.065} \\
$~$  SignCombo & 0.219 & 0.146 &   1.493 &               0.094 &         0.218 &                  7.5\% &    2.316 &   1.061 &    55.2\% &                 1.036 \\
\bottomrule
\end{tabular}
    \begin{tablenotes}
      \item[a] Best performance is in bold. There is no comparison for vol. and downside deviation for Panel B.
      \item[b] Two mixed signals of LinReg and GMOM are included for the purpose of diversification analysis, see Section \ref{sec:backtest_diverse}. RegCombo is constructed from a linear regression with 8 momnetum features (for LinReg) and 8 network momentum features (for GMOM). SignCombo is 50\% LinReg signals plus 50\% GMOM signals. 
    \end{tablenotes}
\end{threeparttable}
\end{table*}

Based on the figures in Table \ref{tab:bt_perf_table_combined}, the Long Only strategy, which serves as the market benchmark, shows modest performance across the board. The model-free baseline, MACD, is slightly less efficient than the Long Only strategy in terms of most metrics. Compared to the model-free MACD strategy, LinReg shows superior performance in terms of both profitability and risk management. It not only has higher returns in both raw signals (0.041 vs 0.025 in Panel A) and volatility-scaled signals (0.165 vs 0.103 in Panel B), but also a better Sharpe ratio, indicating superior risk-adjusted returns. Furthermore, with lower values for downside deviation and maximum drawdown, LinReg is able to better handle risk characteristics and discover more profitable momentum patterns. We will discuss whether the superiority comes from taking into consideration of reversal effects in Section \ref{sec:robust_analysis_reversal}. \par

Next, we introduce the proposed method GMOM. GMOM considers the interconnected system of individual assets' momentum features and how they influence each other. This strategy effectively identifies patterns missed by individual momentum strategies, including LinReg, despite having the same information exposure. Through this unique approach, GMOM yields superior results. As evidence of its robust performance, GMOM leads in terms of expected return (7\% in raw signals and 22\% after volatility scaling), Sharpe ratio (1.363 in raw signals and 1.511 after volatility scaling), and it notably records the smallest Maximum Drawdown (MDD) and the shortest MDD duration. \par

These impressive metrics can also be observed in the plots of cumulative returns in Figure \ref{fig:cumulative_returns}, where the red curve, representing GMOM, outperforms LinReg, MACD, and Long Only across nearly the entire backtest period from 2000 to 2022. This superior performance is observed not only in raw signals but also in volatility-scaled signals. These results suggest that GMOM not only garners higher returns but also manages risk more effectively, demonstrating the added value of considering the network effects in momentum-based strategies. \par

\subsection{Diversification Analysis}
 \label{sec:backtest_diverse}

\begin{figure}[h!]
    \centering
        \begin{subfigure}[]{0.33\textwidth}
        \includegraphics[width=1\textwidth]{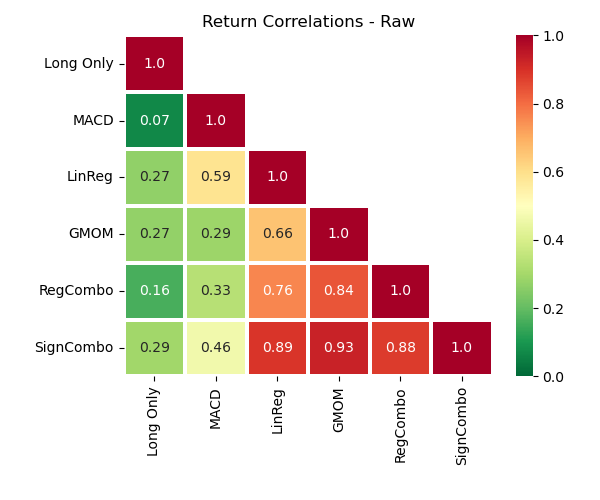}
        \caption{Raw Return Correlations}
        \label{subfig:bt_corr_raw}
    \end{subfigure}%
    \begin{subfigure}[]{0.33\textwidth}
        \includegraphics[width=1\textwidth]{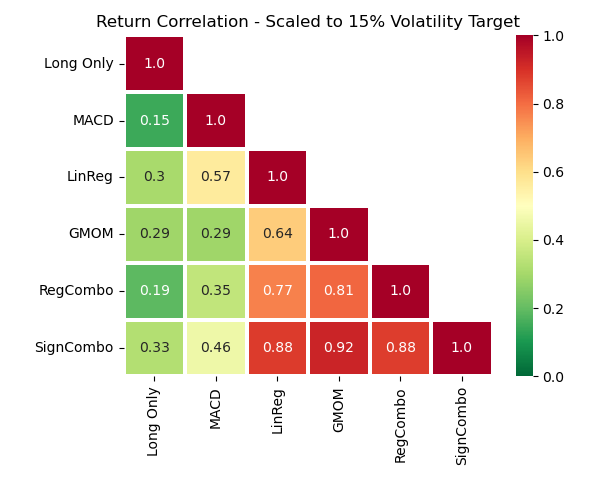}
        \caption{Scaled Return Correlations}
        \label{subfig:bt_corr}
    \end{subfigure}%
    \begin{subfigure}[]{0.33\textwidth}
        \includegraphics[width=1\textwidth]{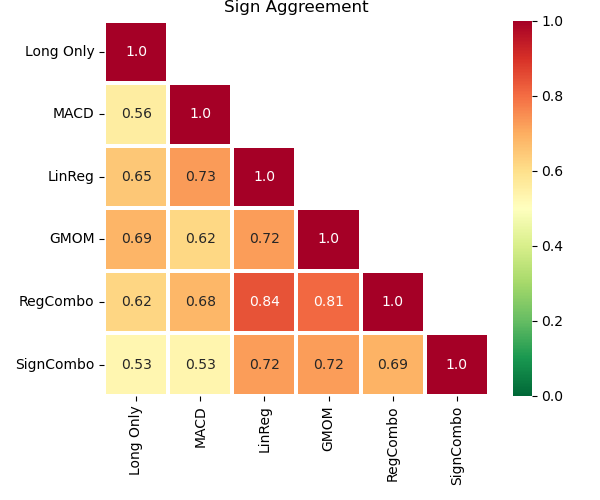}
        \caption{Sign Agreement}
        \label{subfig:sign_agreemtn}
    \end{subfigure}
\caption{A diversification analysis on the correlations and sign agreement of daily returns between the proposed strategy GMOM, three reference strategies (Long Only/MACD/LinReg) and two mixed strategies from LinReg and GMOM (RegCombo/SignCombo) of the entire out-of-sample period from 2000 to 2022. The values of the pairwise correlation and sign agreement are marked in the corresponding square boxes. Sign agreement is defined as the percentage of instances where two signals share the same trading direction for an asset on a trading day, throughout the entire asset universe and backtest period.}
\label{fig:diversification_analysis}
\end{figure}

To examine whether GMOM and LinReg contain orthogonal trading signals, even though they have identical input features, we calculate the correlation of their returns from both raw and volatility-scaled signals in Figure \ref{subfig:bt_corr_raw} and Figure \ref{subfig:bt_corr}. In addition, we evaluate the sign agreement in Figure \ref{subfig:sign_agreemtn}, defined as the percentage of instances when GMOM and LinReg share the same trading direction for an asset on a specific trading day, throughout the entire asset universe and backtest period. \par

We also consider mixing the trading signals of LinReg and GMOM using the following two combo portfolios:
\begin{itemize}
    \setlength\itemsep{-0.5em}
    \item \textbf{RegCombo}: The signals obtained from the regression model that takes both individual momentum and network momentum features as covariates such that  $x_{i,t}^{\text{SignCombo}} = \text{sign}({y}_{i,t}^{\text{RegCombo}})$, where 
    \begin{equation}
    y_{i,t}^{\text{SignCombo}} = \mathbf{u}_{i,t}^{T} \boldsymbol{\beta}_1 +  \tilde{\mathbf{u}}_{i,t}^{T} \boldsymbol{\beta}_2 + b.    
    \end{equation}
    
    \item \textbf{SignCombo}: $x_{i,t}^{\text{SignCombo}} = \frac{1}{2} x_{i,t}^{\text{LinReg}} +  \frac{1}{2} x_{i,t}^{\text{GMOM}}$. 
\end{itemize}

Analysing the correlation of returns between LinReg and GMOM, we find it to be around 65\% for both raw and volatility-scaled signals, and a sign agreement of 72\%. These values imply a level of independence in the trading signals generated by these two strategies. In other words, GMOM seems to capture additional trading signals beyond those identified by LinReg. Despite this, they are not fully orthogonal. When we examine the performance of SignCombo, which takes the average of GMOM and LinReg's positions, we find that it does not surpass GMOM, but it does perform better than LinReg. This suggests that while GMOM appears to cover the trading signals in LinReg, it also offers additional unique signals, resulting in its superior performance.  \par


On the other hand, combining the individual momentum features and network momentum features in a linear regression model (RegCombo) enhances the portfolio's performance beyond GMOM, albeit slightly. This improvement indicates that a better portfolio can be generated when these two types of features are combined. This leads us to an interesting future direction: we have observed these results under a linear model (RegCombo) which uses both individual momentum features and network momentum features. It is plausible that the feature space might be better spanned under nonlinear transformation. Machine learning methods, with their capacity to capture complex, nonlinear relationships, could potentially be employed to further enhance the synergistic combination of individual momentum and network momentum features for superior portfolio performance.  \par

Overall, these results imply that GMOM, with its unique approach to considering network effects in momentum-based strategies, provides additional value beyond individual momentum strategies like LinReg. It is important to note, however, that properly combining these strategies can result in even better performance, suggesting potential further research and development. \par

\subsection{Turnover Analysis}

\begin{figure}[]
    \centering
        \begin{subfigure}[]{0.5\textwidth}
        \includegraphics[width=1\textwidth]{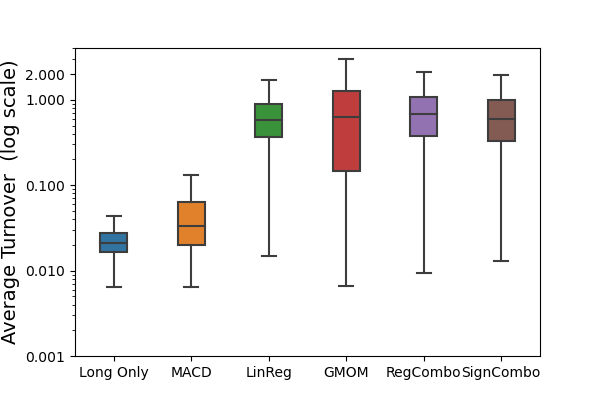}
        \caption{Average Turnover}
        \label{subfig:turnover}
    \end{subfigure}%
    \begin{subfigure}[]{0.5\textwidth}
        \includegraphics[width=1\textwidth]{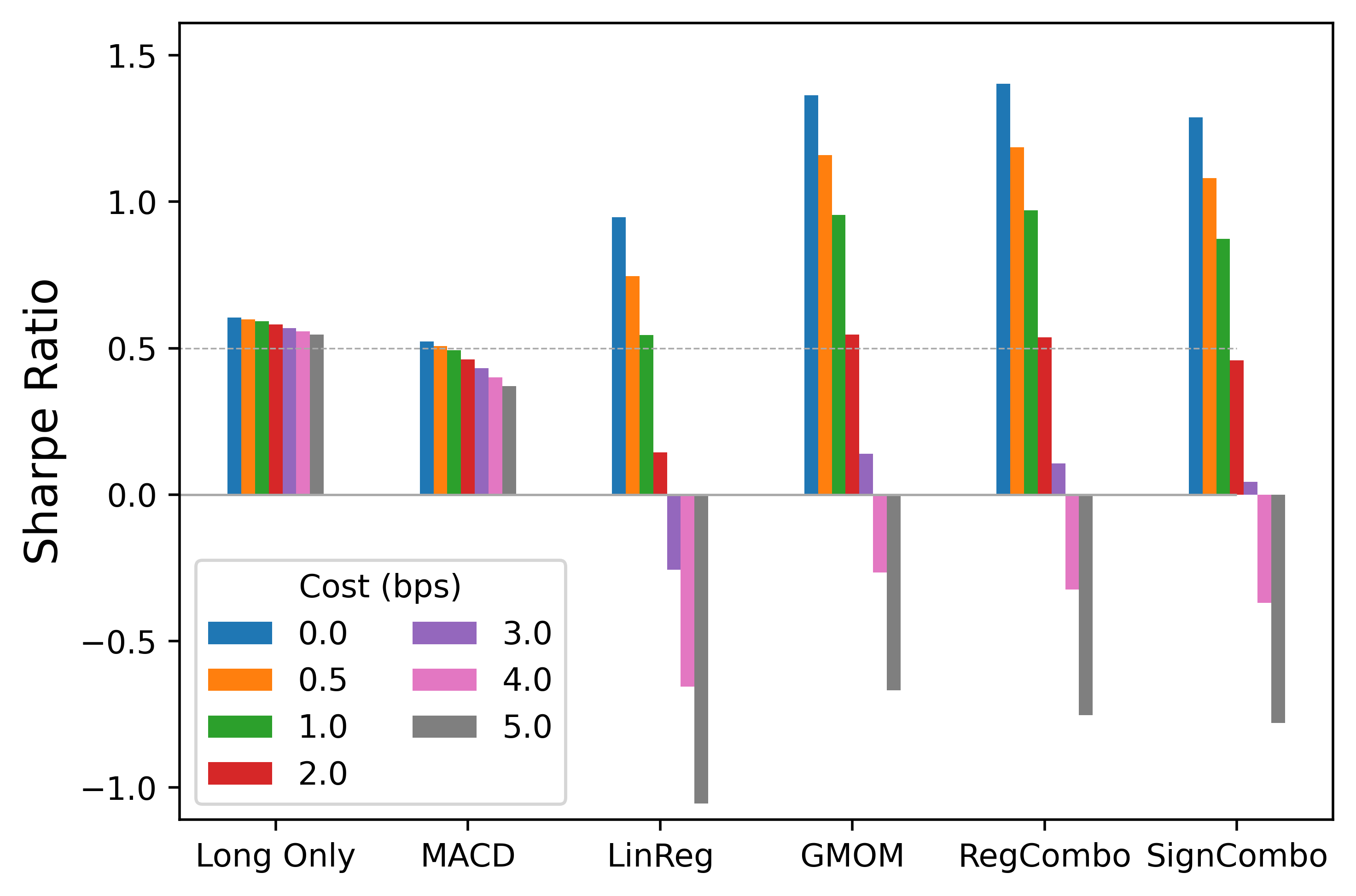}
        \caption{Cost-adjusted Sharpe}
        \label{subfig:sr_turnover}
    \end{subfigure}
\caption{A turnover analysis on the average turnover and cost-adjusted Sharpe ratio between the proposed strategy GMOM, three reference strategies (Long Only/MACD/LinReg) and two mixed strategies from LinReg and GMOM (RegCombo/SignCombo) of the entire out-of-sample period from 2000 to 2022.}
\label{fig:Turnover_analysis}
\end{figure}

In this section, we analyse how the network momentum strategy performs under transaction costs. Following the convention in literature \cite{limEnhancingTimeSeries2020}, we define the turnover $\zeta_{i,t}$ to describe the daily position changes of signals for asset $i$ from trading day $t-1$ to day $t$ such that
\begin{equation}
    \zeta_{i,t} = \sigma_{\text{tgt}} \left|\frac{x_{i,t}}{\sigma_{i,t}}- \frac{x_{i,t-1}}{\sigma_{i,t-1}} \right|
\end{equation}%
We plot the average turnover $\zeta_{i,t}$ of all the assets in our universe across the whole backtest period in Figure \ref{subfig:turnover}. In figure \ref{subfig:sr_turnover}, 
we also present the cost-adjusted Sharpe ratio of strategies, which is calculated from the cost-adjusted returns defined as
\begin{equation}
    \Tilde{r}_{t:t+1}^{\text{ portfolio}} = \frac{1}{N_t}\sum_{i=1}^{N_t}\left(x_{i,t} \frac{\sigma_{\text{tgt}}}{\sigma_{i,t}}r_{i, t:t+1} - c \cdot \zeta_{i,t} \right) 
\end{equation}%
where $c$ is the pseudo-cost in basis points (bps). We consider $c = \{0, 0.5, 1, 2, 3, 4, 5\}$ bps in Figure \ref{subfig:turnover}. \par

The turnover of the strategies illustrates that model-based methods, such as LinReg, GMOM, RegCombo, and SignCombo, exhibit higher turnovers, which is consistent with findings in the literature \cite{limEnhancingTimeSeries2020}. These methods frequently update their market positions to incorporate new information and adapt to market conditions. Conversely, model-free methods like the MACD strategy, which employ fixed trading rules, have a lower turnover of 0.058. The Long Only strategy, which maintains a constant position, records the lowest turnover at 0.025, occurring only when assets cease trading on certain days. \citet{limEnhancingTimeSeries2020} proposed turnover regularisation in model-based momentum strategies, which significantly improved turnover. We reserve this for future research. \par

In a scenario of no transaction costs, LinReg and GMOM significantly surpass the Long Only and MACD strategies, with Sharpe ratios of 0.947 and 1.363, respectively. However, their performance declines as transaction costs rise. At a cost level of 5bps, LinReg's performance significantly reduces to -1.05, while GMOM maintains a slightly better, though still negative, cost-adjusted Sharpe ratio of -0.67. Interestingly, GMOM demonstrates resilience to transaction costs. It manages to maintain a positive Sharpe ratio up to a cost of 3bps, and a ratio above 0.5 up to 2bps. This aligns with practical trading requirements, suggesting that GMOM is more adept at handling transaction costs than LinReg. \par

\section{Robustness Analysis}

\subsection{Graph Topology Analysis}

\begin{figure*}[h!]
\centering
    \begin{subfigure}[]{0.49\textwidth}
        \includegraphics[width=1\textwidth]{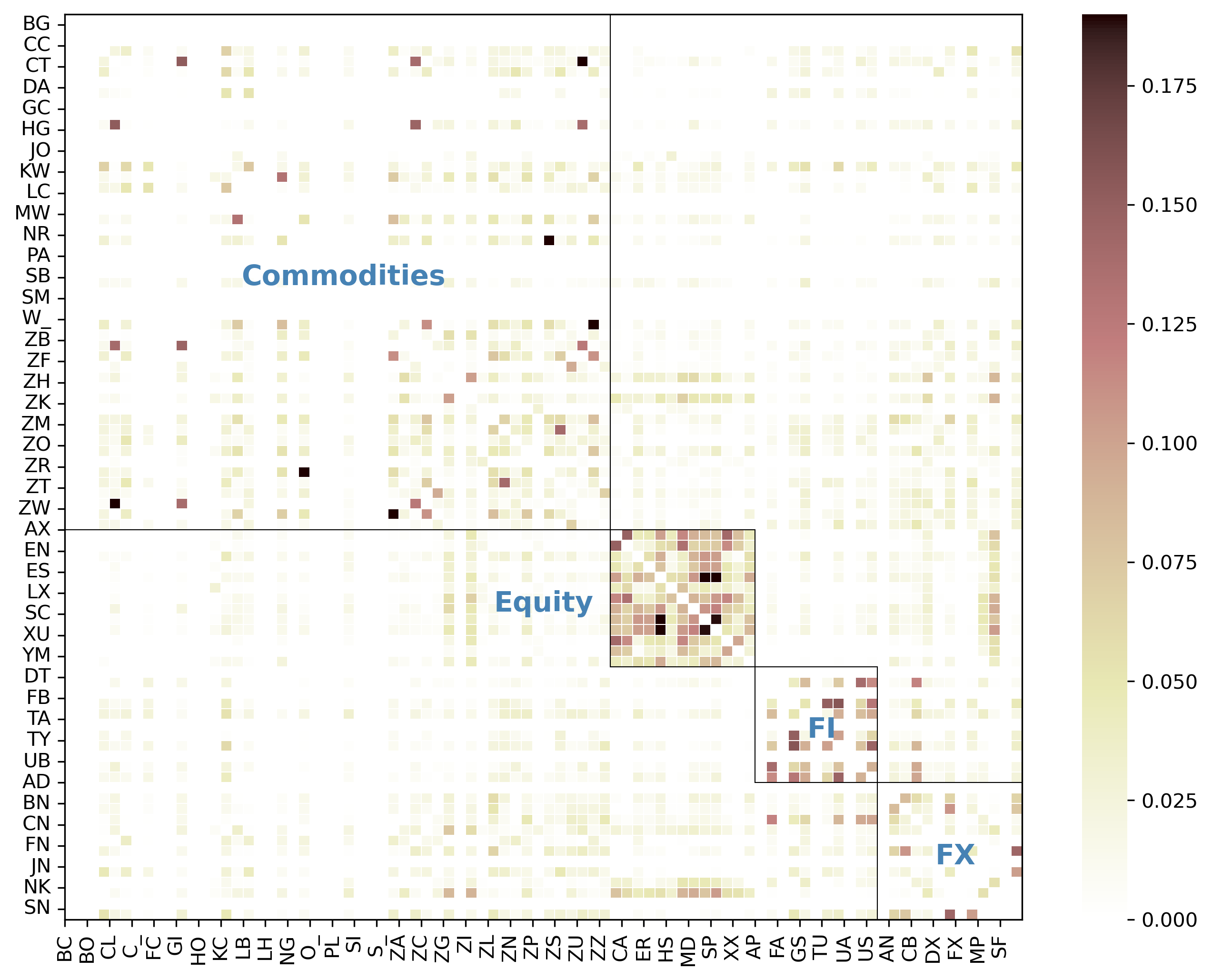}
        \caption{2006-12-22}
        \label{subfig:graph_gmom_2006}
    \end{subfigure}%
    \begin{subfigure}[]{0.49\textwidth}
        \includegraphics[width=1\textwidth]{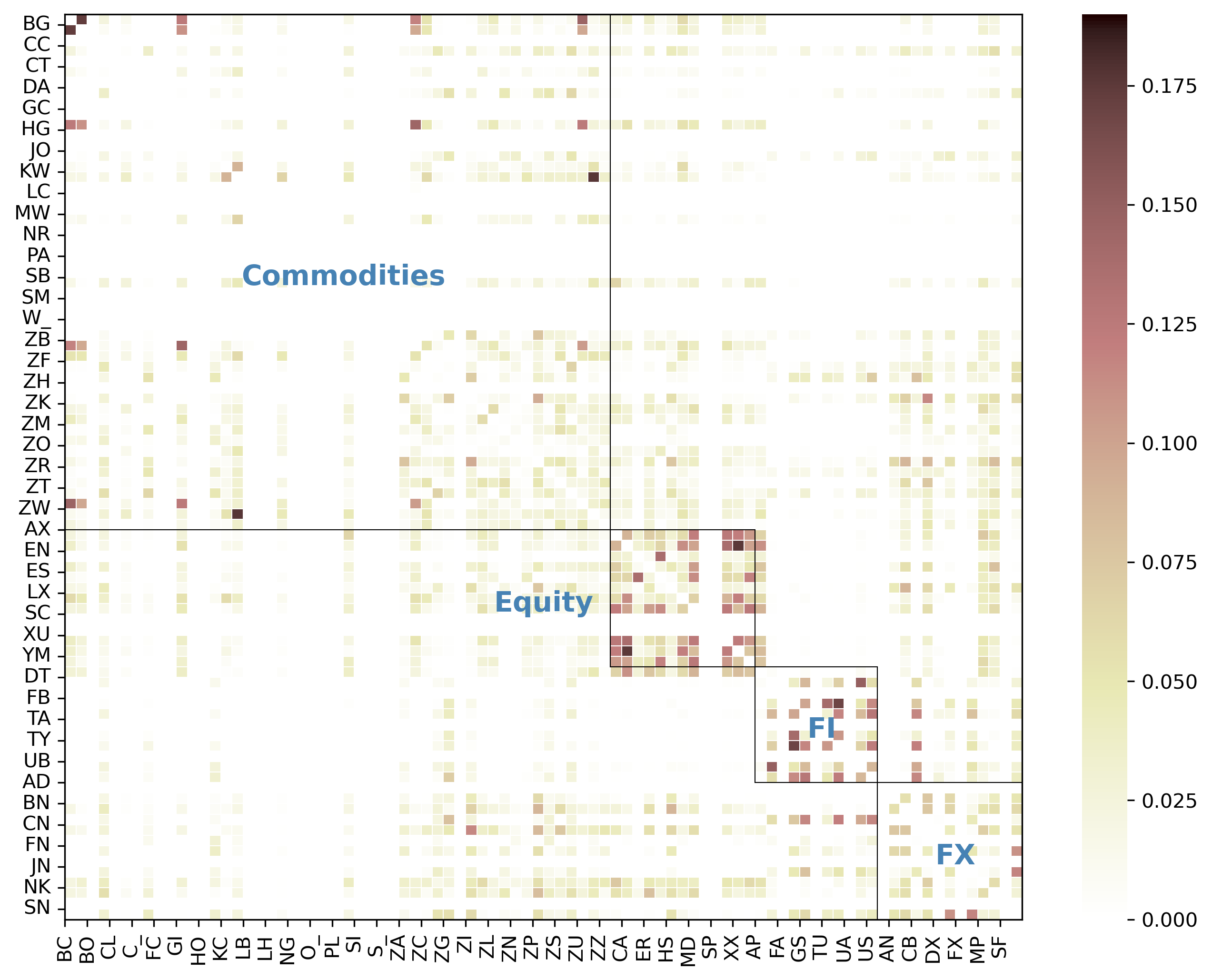}
        \caption{2022-04-05}
        \label{subfig:graph_gmom_2022}
    \end{subfigure}
\caption{Networks obtained from graph learning. Note that the network is an ensemble result of 5 lookback windows spanning the previous five years. The assets are sorted according to their asset classes.}
\label{fig:heatmap_4examples}
\end{figure*}

\begin{figure*}[h!]
\centering
    \begin{subfigure}[]{0.5\textwidth}
        \includegraphics[width=1\textwidth]{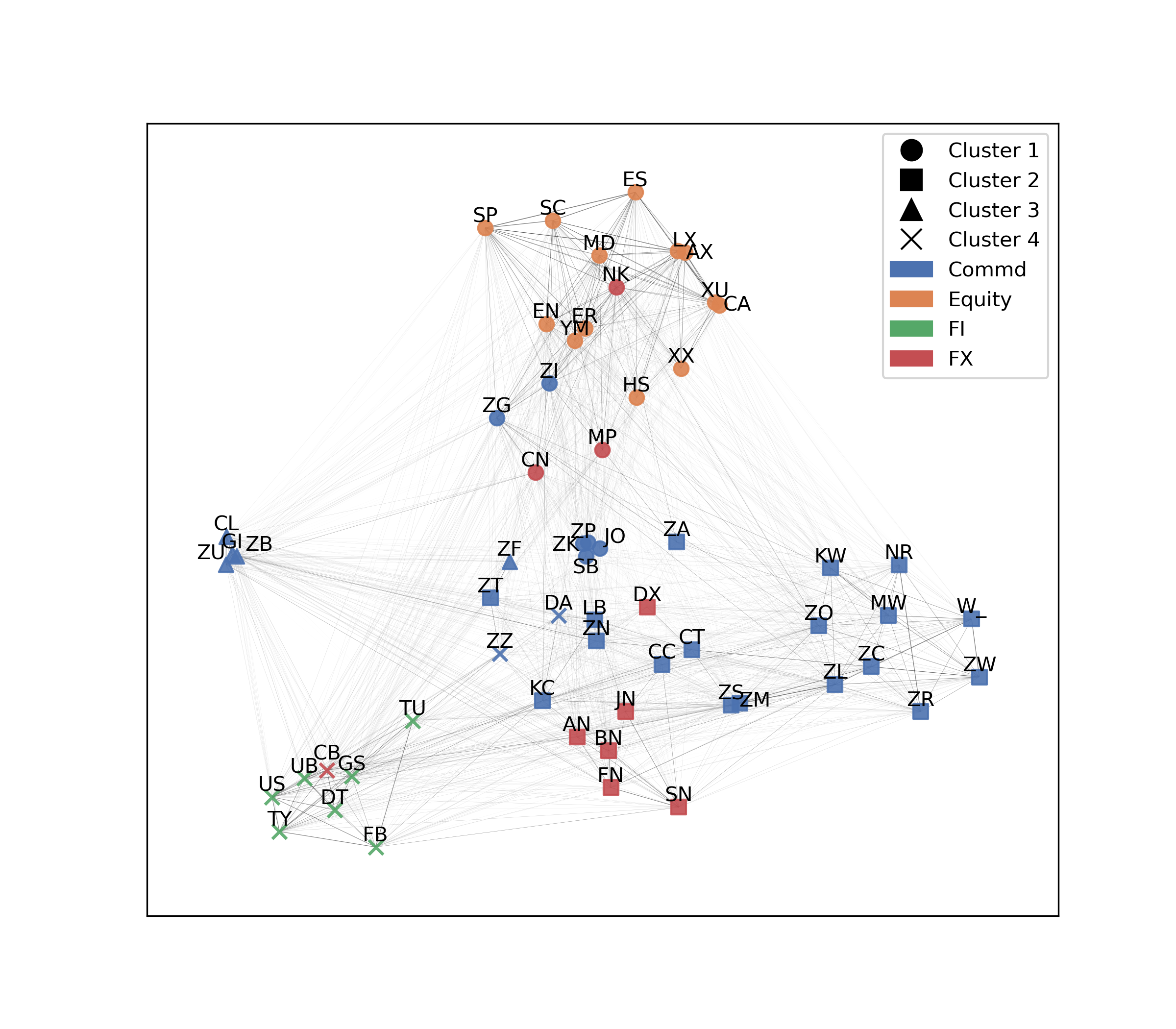}
        \caption{2006-12-22}
        \label{subfig:tsnegraph_gmom_2006}
    \end{subfigure}%
    \begin{subfigure}[]{0.5\textwidth}
        \includegraphics[width=1\textwidth]{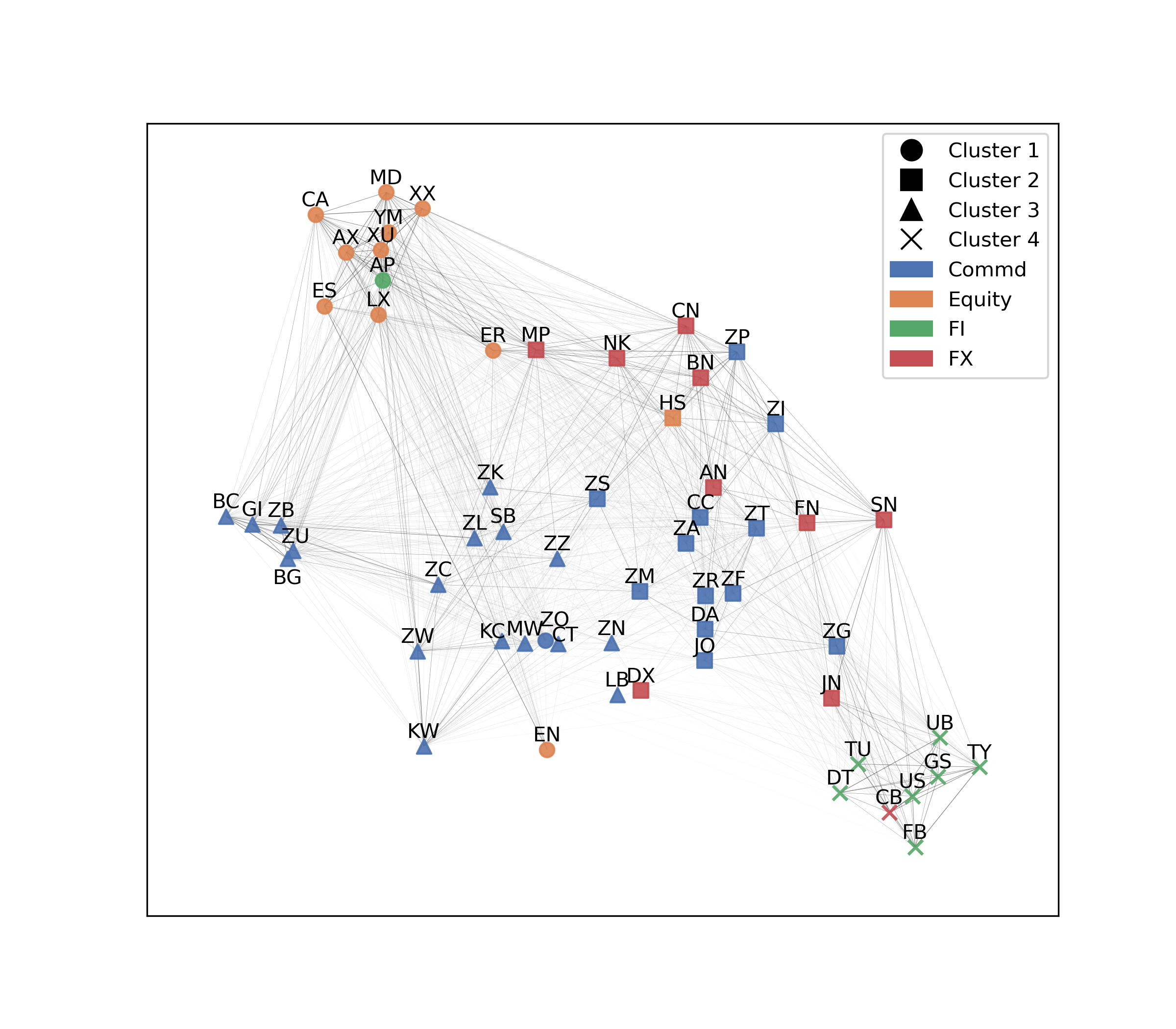}
        \caption{2022-04-05}
        \label{subfig:tsne_graph_gmom_2022}
    \end{subfigure}
\caption{A clustering analysis. Nodes are positioned by t-SNE based on spectral embeddings for spectral clustering and coloured by asset class. Four different node shapes represent four clusters obtained from spectral clustering.}
\label{fig:tsne_4examples}
\end{figure*}

\begin{figure*}[h!]
\centering
    \begin{subfigure}[]{0.33\textwidth}
        \includegraphics[width=1\textwidth]{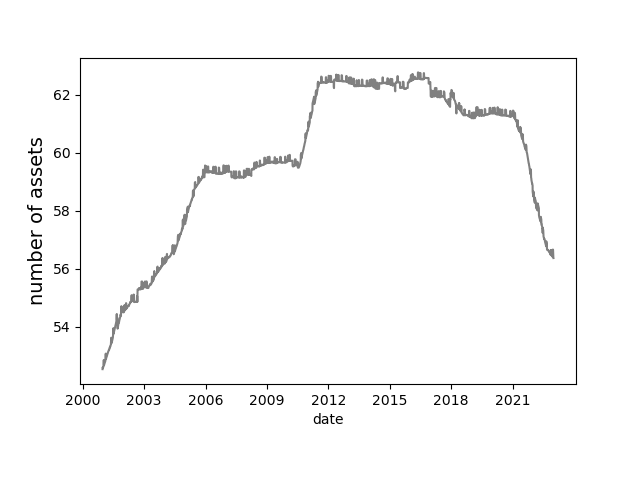}
        \caption{number of nodes (assets)}
        \label{subfig:num_nodes}
    \end{subfigure}%
    \begin{subfigure}[]{0.33\textwidth}
        \includegraphics[width=1\textwidth]{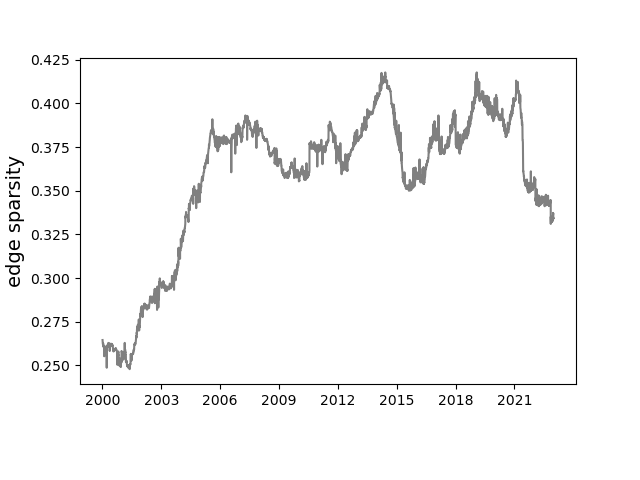}
        \caption{edge sparsity}
        \label{subfig:edge_sparsity}
    \end{subfigure}%
    \begin{subfigure}[]{0.33\textwidth}
        \includegraphics[width=1\textwidth]{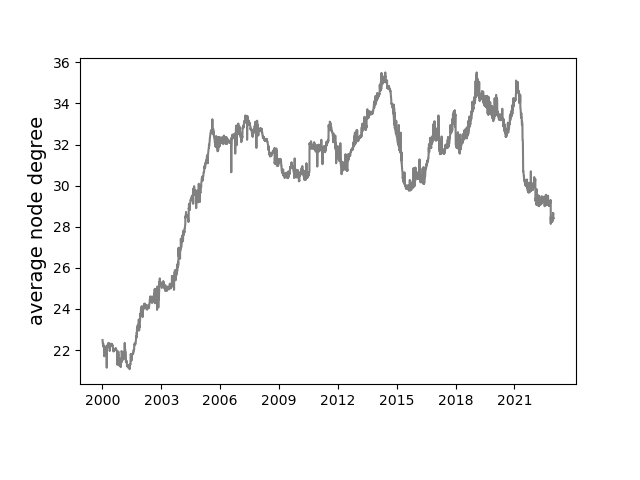}
        \caption{node degree}
        \label{subfig:node_deg_mean}
    \end{subfigure}
    
    \begin{subfigure}[]{0.33\textwidth}
        \includegraphics[width=1\textwidth]{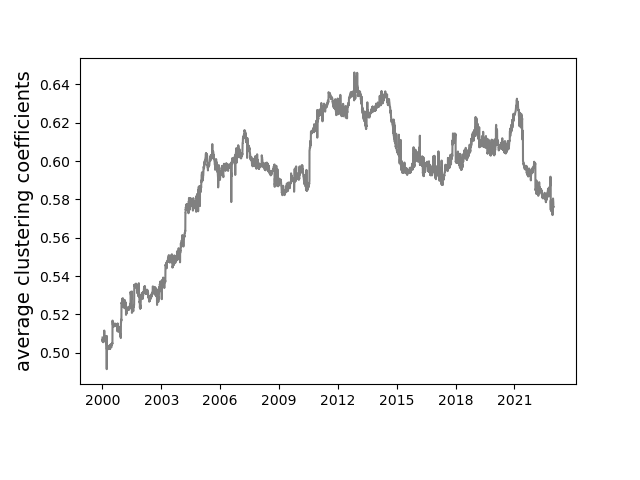}
        \caption{clustering coefficients}
        \label{subfig:avg_clustering}
    \end{subfigure}%
        \begin{subfigure}[]{0.33\textwidth}
        \includegraphics[width=1\textwidth]{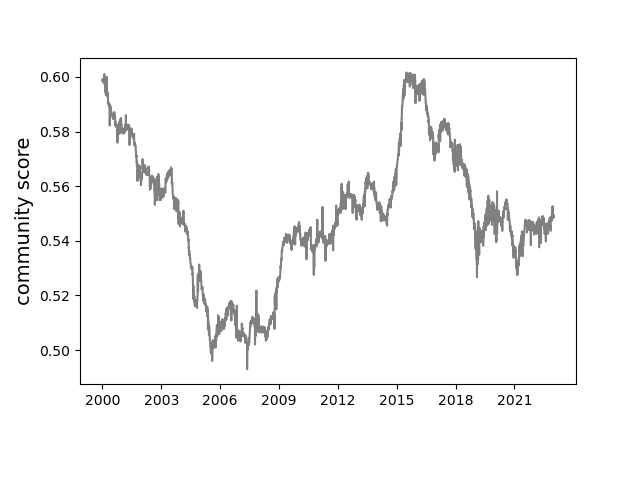}
        \caption{community ratio}
        \label{subfig:community_ratio}
    \end{subfigure}%
        \begin{subfigure}[]{0.33\textwidth}
        \includegraphics[width=1\textwidth]{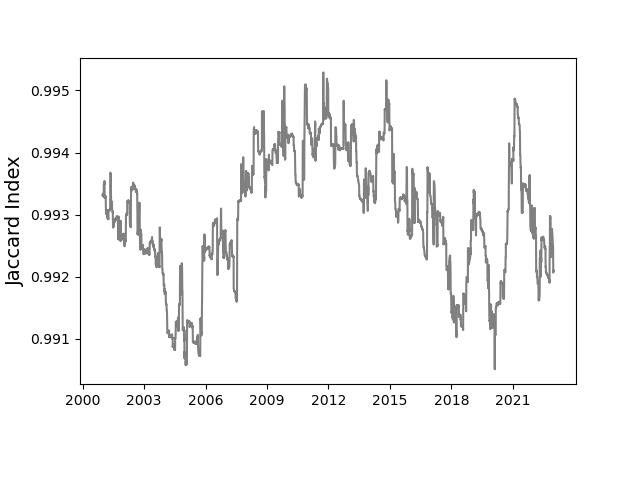}
        \caption{Jaccard index}
        \label{subfig:jaccard}
    \end{subfigure}
\caption{Network statistics for topological analysis across test periods. Note that the network is an ensemble result of 5 lookback windows spanning the previous five years.}
\label{fig:network-stats}
\end{figure*}

In this section, we analyse the topological characteristics of graphs learned for each trading day. We begin by showcasing examples of these learned graphs from a stable period (2006-12-22) and a volatile period (2022-04-05), as illustrated in Figure \ref{fig:heatmap_4examples}. Note that our graph is an ensemble representation of information spanning the previous five years. Consequently, the graph for 2022-04-05 contains the information of 2022 Russian-Ukraine war and 2020 pandemic period. To delve deeper into the community structure and ascertain whether it aligns with the asset classes, we execute spectral clustering on each graph. The clusters, represented by different node shapes in Figure \ref{fig:tsne_4examples}, are colour-coded by asset classes. Node positions are determined by the t-SNE of their spectral embedding. Lastly, we construct a time series for each of these properties. These time series are depicted in Figure \ref{fig:network-stats}. The properties under consideration include:

\begin{itemize}
    \setlength\itemsep{-0.5em}
    \item Number of nodes (assets): $N_t$
    
    \item Edge Sparsity \cite{bornerNetworkScience2007}: $G^{~\text{sparse}}_t = \frac{2 |\mathcal{E}^{(t)}|}{N_t(N_t - 1)}$, where $|\mathcal{E}^{(t)}|$ is the number of edges and $N_t(N_t - 1)$ is the total number of possible edges if every pair of assets would be connected. This metric quantifies the density of a graph. A larger value indicates a denser graph, implying rapid information propagation across the network.
    
    \item Average Node Degree \cite{barabasiNetworkScience2015}: $d_t = \frac{1}{N_t} \sum_i d_{i,t}$, where $d_{i,t}= \sum_{j}^{N_t} \mathbf{1}_{\{ \mathbf{A}_{ij,t}\neq 0\}}$. This is a measure of the average number of connections a node has within the network.
 
    \item Clustering coefficients \cite{kaiserMeanClusteringCoefficients2008}: $G^{\text{ cluster}}_t = \frac{1}{N_t}\sum_{i}^{N_t} \frac{2 T_{i,t}}{d_{i,t} (d_{i,t}-1)}$, where $T_{i,t}$ represents the number of triangles involving node $i$. A triangle refers to a set of three nodes where each node is connected to the other two. A value close to 1 indicates a highly interconnected network with a tendency to form tightly-knit communities, while a value close to 0 suggests a more loosely or randomly connected network.
 
    \item Community ratio \cite{doreianGeneralizedBlockmodeling2004}: $G^{\text{ community}}_t = \frac{ 2 \left( \sum_{i,j} \mathbf{1}_{\{\mathbf{A}_{ij,t} \neq 0 \land C_i = C_j\}} + \sum_{i,j} \mathbf{1}_{\{\mathbf{A}_{ij,t} = 0 \land C_i \neq C_j\}} \right)}{N_t (N_t -1)}$, where $C_i$ is the asset class of asset $i$. This metric measures the prevalence of intra-class edges versus inter-class edges.
 
    \item Jaccard index \cite{zhangGraphbasedMethodsForecasting2022}: $G^{\text{ Jaccard}}_t = \frac{\left|\mathcal{E}^{(t)} \cap \mathcal{E}^{(t-1)} \right|}{\left|\mathcal{E}^{(t)} \cup \mathcal{E}^{(t-1)} \right|}$, where $\mathcal{E}^{(t)} = \{(i,j): \mathbf{A}{ij,t}\neq 0, i \neq j\}$, is used to measure the similarity of edges captured by two consecutive graphs.
 
\end{itemize}

The networks exhibit remarkable stability in their properties. This is evidenced by a consistently high Jaccard index, generally above 0.99, indicating a predominant capture of the same edges over time. The stability extends to the edge sparsity, node degree, and clustering coefficient. It is important to note that these properties are dependent on the number of nodes, and their trends mirror the trends of the number of nodes. The number of nodes, in turn, is influenced by the available assets, which have varying trading days and periods in our database (see Table \ref{table:universe}). The small edge sparsity, a generally value less than 0.4, suggests that the propagation of momentum information would encounter less spread of noise, a desirable characteristic for constructing network momentum. \par

However, the community ratios that measure the prevalence of intra-class edges versus inter-class edges, although remaining at a high level (greater than 0.5) the whole test period, exhibit greater volatility over time. This is particularly noticeable during the global financial crisis of 2007-2008, when the community score drops significantly. This suggests a potential disruption in the intra-class and inter-class connections within the network. Moreover, the Jaccard index, which measures the similarity of edges captured by two consecutive graphs, shows substantial fluctuations during periods of financial turbulence, such as the 2007-2008 financial crisis and the 2020 pandemic. These fluctuations may reflect the dynamic restructuring of the network in response to external shocks, underscoring the sensitivity of the network topology to macroeconomic conditions. \par


The node degree remains relatively stable over time, indicating that each asset maintains a consistent number of connections within the network. Despite the volatile trends in community ratios, the learned graph exhibits some community structure corresponding to asset classes, with the value exceeding 0.5 for most of the backtest period. By comparing the clusters obtained from spectral clustering to asset classes, as shown in Figure \ref{fig:tsne_4examples}, many assets that belong to the same asset class are still in the same cluster. However, there are instances where assets from different asset classes are clustered into the same group. For instance, on 2006-12-22, Milk (DA) and Lean Hogs (ZZ) from commodities are clustered into a FI-dominant cluster. It is noteworthy that CANADIAN 10YR BOND (CB) was incorrectly categorised into FX in our raw database \cite{limEnhancingTimeSeries2020}. We retained this mistake to examine whether our methods could reveal that CB aligns more closely with FI, a hypothesis confirmed by results from both 2006-12-22 and 2022-04-05. \par

Interestingly, FI assets form tight clusters, with the exception of Australian price index (AP), which has strong connections with all the assets in Equity. This can be observed from the heatmaps in Figure \ref{fig:heatmap_4examples} and in Figure \ref{fig:tsne_4examples}. It should be noted that AP only has pricing data from 2010. \par

Some FX form a cluster with commodities, while other FX form a cluster with equities, indicating inter-class similarity in momentum features. By comparing 2006-12-22 (a calm period) and 2022-04-05 (a turbulent period), we observe that in the calm period, there are still connections between equity and FI, while in the turbulent period, such edges are absent. During turbulent times, equities form an even denser cluster, with only AP in the cluster, as shown in Figure \ref{fig:tsne_4examples}. However, in 2006-12-22, equities, along with some commodities (GOLD/SILVER/PLatinum/copper/orange juice/sugar - ZG/ZI/ZP/ZK/JO/SB) and some FX assets (NK/CN/MP), form a cluster.

These observations suggest that the community structure does not align perfectly with asset classes. There are many inter-class assets in equities, FX, and commodities that exhibit strong similarity in momentum features.

\subsection{Individual Asset Class Portfolios}
\label{sec:individual-asset-class-portfolio}
\begin{table}[h!]
\centering
\small
\caption{Performance of GMOM portfolios across asset classes from signals rescaled to 15\% volatility target}
\label{tab:robust_perf_table_scaled}

\begin{threeparttable}
\begin{tabular}{lcccccccccc}
\toprule
{} &    return &    vol. &  Sharpe &  \makecell{downside \\ deviation} & \makecell{MDD} &  \makecell{MDD \\ duration} &  Sortino &  Calmar &  hit rate &  $\frac{\text{Avg. P}}{\text{Avg. L}}$ \\
\midrule
GMOM      & \bf{0.222} & 0.147 &   \bf{1.511} &               0.092 &         \bf{0.199} &                  \bf{6.9\%} &    \bf{2.422} &   \bf{1.179} &    \bf{55.2\%} &                 1.038 \\

\midrule
\multicolumn{11}{c}{Panel A: Multi-class GMOM with Intra- or Inter-Class Edges}  \\
\midrule
GMOM-Intra & 0.177 & 0.147 &   1.206 &               0.094 &         0.293 &                  9.3\% &    1.892 &   0.616 &    54.2\% &                 1.026 \\
GMOM-Inter & 0.134 & 0.147 &   0.911 &               0.094 &         0.607 &                 27.3\% &    1.429 &   0.215 &    53.5\% &                 1.009 \\

\midrule
\multicolumn{11}{c}{Panel B: Individual Asset Class Portfolio from Cross-class Network Momentum} \\
\midrule
M-Commd & 0.131 & 0.146 &   0.895 &               0.094 &         0.438 &                 14.9\% &    1.388 &   0.291 &    53.1\% &                 1.021 \\
M-Equity & 0.138 & 0.146 &   0.948 &               0.099 &         0.274 &                 16.5\% &    1.403 &   0.497 &    54.0\% &                 0.995 \\
M-FI     & 0.180 & 0.147 &   1.225 &               0.092 &         0.396 &                 13.3\% &    1.962 &   0.466 &    53.7\% &                 \bf{1.050} \\
M-FX    & 0.097 & 0.147 &   0.657 &               0.093 &         0.572 &                 28.7\% &    1.039 &   0.157 &    51.8\% &                 1.033 \\

\midrule
\multicolumn{11}{c}{Panel C: Individual Asset Class Portfolio from Single-class Network Momentum}\\
\midrule
S-Commd    & 0.117 & 0.145 &   0.807 &               0.095 &         0.399 &                 15.6\% &    1.234 &   0.283 &    52.8\% &                 1.022 \\
S-Equity    & 0.044 & 0.146 &   0.302 &               0.096 &         0.777 &                 46.0\% &    0.459 &   0.044 &    52.1\% &                 0.965 \\
S-FI        & 0.115 & 0.147 &   0.781 &               0.089 &         0.389 &                 12.1\% &    1.290 &   0.282 &    51.9\% &                 1.052 \\
S-FX        & 0.024 & 0.147 &   0.162 &               0.094 &         0.701 &                 40.2\% &    0.252 &   0.019 &    50.9\% &                 0.989 \\
\bottomrule
\end{tabular}
\begin{tablenotes}
  \item[a] Cross-panel best performance is in bold. No comparison of vol. and downside deviation for volatility-scaled signals.
  \item[b] Four classes include Commodities (Commd), Equities (Equity), Fixed Income (FI) and Foreign Currencies (FX)
  \item[c] For the construction details of every portfolio, refer to Section \ref{sec:individual-asset-class-portfolio}.
\end{tablenotes}
\end{threeparttable}
\end{table}

\begin{figure}[h!]
\centering
    \begin{subfigure}[]{0.333\textwidth}
        \includegraphics[width=1\textwidth]{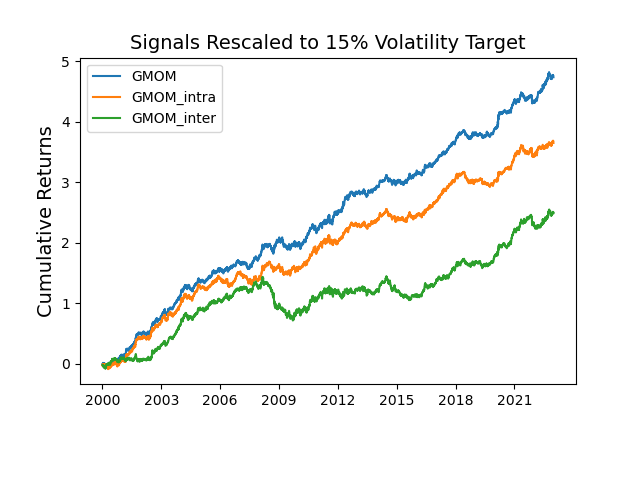}
        \caption{Panel A Portfolios}
        \label{subfig:PanelA_port_pnl}
    \end{subfigure}%
    \begin{subfigure}[]{0.333\textwidth}
        \includegraphics[width=1\textwidth]{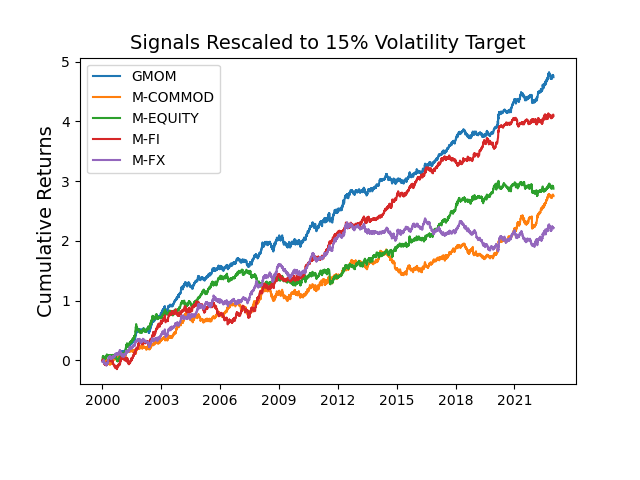}
       \caption{Panel B Portfolios}
        \label{subfig:PanelB_port_pnl}
    \end{subfigure}%
    \begin{subfigure}[]{0.333\textwidth}
        \includegraphics[width=1\textwidth]{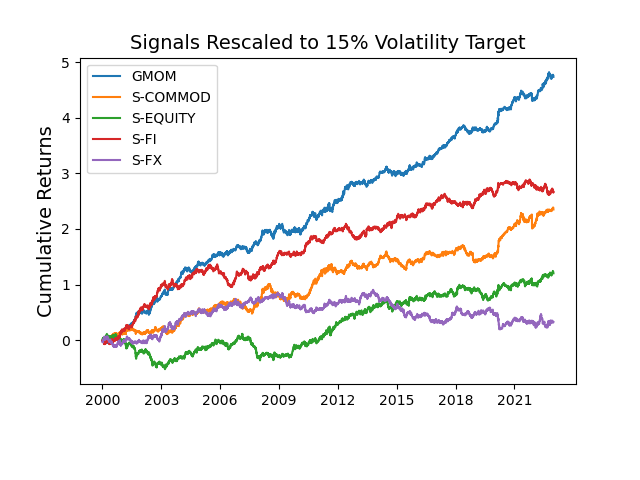}
        \caption{Panel C Portfolios}
        \label{subfig:PanelC_port_pnl}
    \end{subfigure}
\caption{The cumulative daily returns of the volatility-scaled signals of the individual asset class portfolios portfolios displayed in Table \ref{tab:robust_perf_table_scaled}.}
\label{fig:class_port_pnl}
\end{figure}

\begin{figure}[h!]
\centering
    \begin{subfigure}[]{0.333\textwidth}
        \includegraphics[width=1\textwidth]{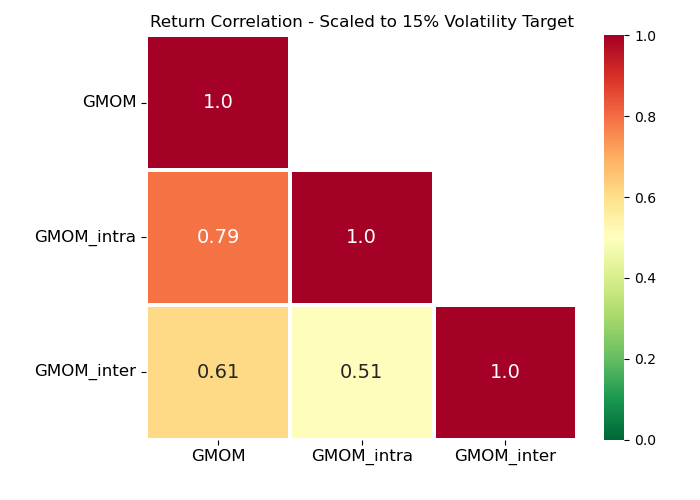}
        \caption{Panel A Portfolios}
        \label{subfig:PanelA_port_corr}
    \end{subfigure}%
    \begin{subfigure}[]{0.333\textwidth}
        \includegraphics[width=1\textwidth]{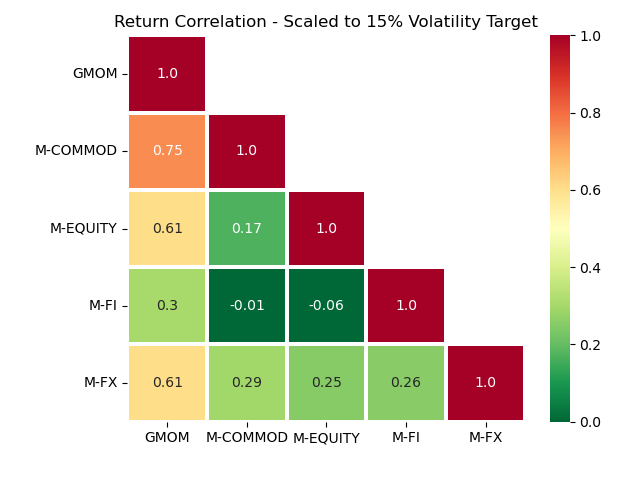}
       \caption{Panel B Portfolios}
        \label{subfig:PanelB_port_corr}
    \end{subfigure}%
    \begin{subfigure}[]{0.333\textwidth}
        \includegraphics[width=1\textwidth]{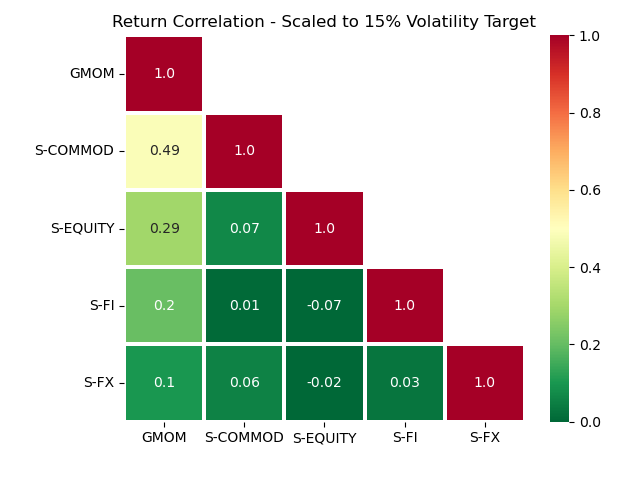}
        \caption{Panel C Portfolios}
        \label{subfig:PanelC_port_corr}
    \end{subfigure}
\caption{The return correlations of volatility-scaled signals of the individual asset class portfolios portfolios displayed in Table \ref{tab:robust_perf_table_scaled}.}
\label{fig:class_port_corr}
\end{figure}

Based on the primary exhibits from Section \ref{sec:4backtest}, GMOM displayed superior profitability and effectiveness in mitigating risk with lower volatility and downside durations. This reflects the advantage of incorporating network effects learned from past individual momentum for alpha generation and portfolio construction. This section delves deeper into the robustness of this performance. Specifically, we investigate whether the observed performance is an outcome of the network effect across multiple asset classes or it hinges predominantly on a single asset class. \par

To address this, we construct several ablation portfolios and report their performance in Table \ref{tab:robust_perf_table_scaled} for volatility-rescaled signals. These portfolios are designed as follows:
\begin{itemize}
    \setlength\itemsep{-0.5em}
    \item \textbf{Panel A} - portfolios that selectively deactivate inter-class or intra-class edges in the learned graphs and then construct network momentum respectively. GMOM-Intra only considers intra-class edges, while GMOM-Inter only consider inter-class edges. This allows us to understand whether the performance of GMOM relies more on the internal connections within an asset class (intra-class edges) or the connections between different asset classes (inter-class edges).
    
    \item \textbf{Panel B} - portfolios that trades only on assets of a single class, based on the same network momentum signals from the cross-asset class networks as in GMOM. The purpose is to determine if the advantages of GMOM's network momentum persist when its applicability is limited to a single asset class.
    
    \item \textbf{Panel C} - portfolios that follow our methodology to construct graphs for each asset class individually, calculating the network momentum within these distinct classes. This approach allows us to evaluate whether the network momentum derived from individual asset classes alone can deliver performance comparable to GMOM, by isolating the network effects within individual asset classes.
\end{itemize}%
We also plot the cumulative returns and correlations of daily returns from the volatility-rescaled signals for comparing the portfolio of each panel in Figure \ref{fig:class_port_pnl} and Figure \ref{fig:class_port_corr}. \par

From Panel A in Table \ref{tab:robust_perf_table_scaled}, both GMOM-Intra and GMOM-Inter report inferior performance metrics compared to the original GMOM portfolio. The intra-class version (GMOM-Intra) appears to perform better than the inter-class (GMOM-Inter), indicating that connections within an asset class contribute more to GMOM's overall performance than those between different asset classes. Despite this, a portfolio with a favourable Sharpe ratio can still be constructed solely from the inter-class connections, thereby acknowledging their value. This is also reflected in their cumulative returns (Figure \ref{subfig:PanelA_port_pnl}), where GMOM-Inter experienced a significant drop during the global financial crisis (2007-2008), demonstrating that assets from different classes behave differently in turbulent times. However, due to their low correlation (0.51), GMOM, as a combination of these two types of edges, can have superior portfolio performance. The inter-class connections also lead to a larger maximum drawdown and longer drawdown durations, underscoring the importance of intra-class network effects in buffering downside risk and maintaining portfolio stability. \par

By analysing Panel B, while Fixed Income (M-FI) delivers robust performance in isolation, it may not be the dominant contributor to GMOM's performance due to its relatively low correlation with GMOM (0.3). In contrast, Commodities (M-Commd), despite having a lower standalone return, demonstrates a higher correlation with GMOM (0.75). Note that this is in part also related to the larger number of commodity futures overall. Given that we probably have most commodities assets, this indicates a significant contribution to GMOM's overall performance. This distinction highlights the complexity of the cross-asset class interplay and the importance of network momentum signals in influencing GMOM's performance. Each asset class, while maintaining strong performance in isolation, also contributes to a well-diversified portfolio due to low cross-correlations, thereby reinforcing the value of GMOM's cross-asset class approach. \par

Comparing Panel B and Panel C in Table \ref{tab:robust_perf_table_scaled}, it becomes evident that the performance of portfolios that are restricted to single asset classes (Panel B), yet still leveraging network momentum signals from cross-class connections, generally surpasses those constructed from the network momentum within individual asset classes (Panel C). This superiority demonstrates the effectiveness of the network momentum derived from cross-asset class connections. Take the Fixed Income (FI) asset class as an example. In Panel B (multi-class GMOM applied to individual asset classes), the return is 0.180, the Sharpe ratio is 1.225, and the Calmar ratio is 0.466. In contrast, in Panel C (single-class GMOM applied to individual asset classes), the return decreases to 0.115, the Sharpe ratio to 0.781, and the Calmar ratio to 0.282. Similar trends are observed in all other three asset classes. This comparison highlights the significance of cross-asset class network effects. By capturing inter-asset class momentum, they provide valuable portfolio management information that is not accessible when focusing on individual asset classes. \par

\subsection{Graph Learning with Different Lookback Windows}
\label{sec:robust_lookback}

\begin{figure*}[h!]
\centering
    \begin{subfigure}[]{0.195\textwidth}
        \includegraphics[width=1\textwidth]{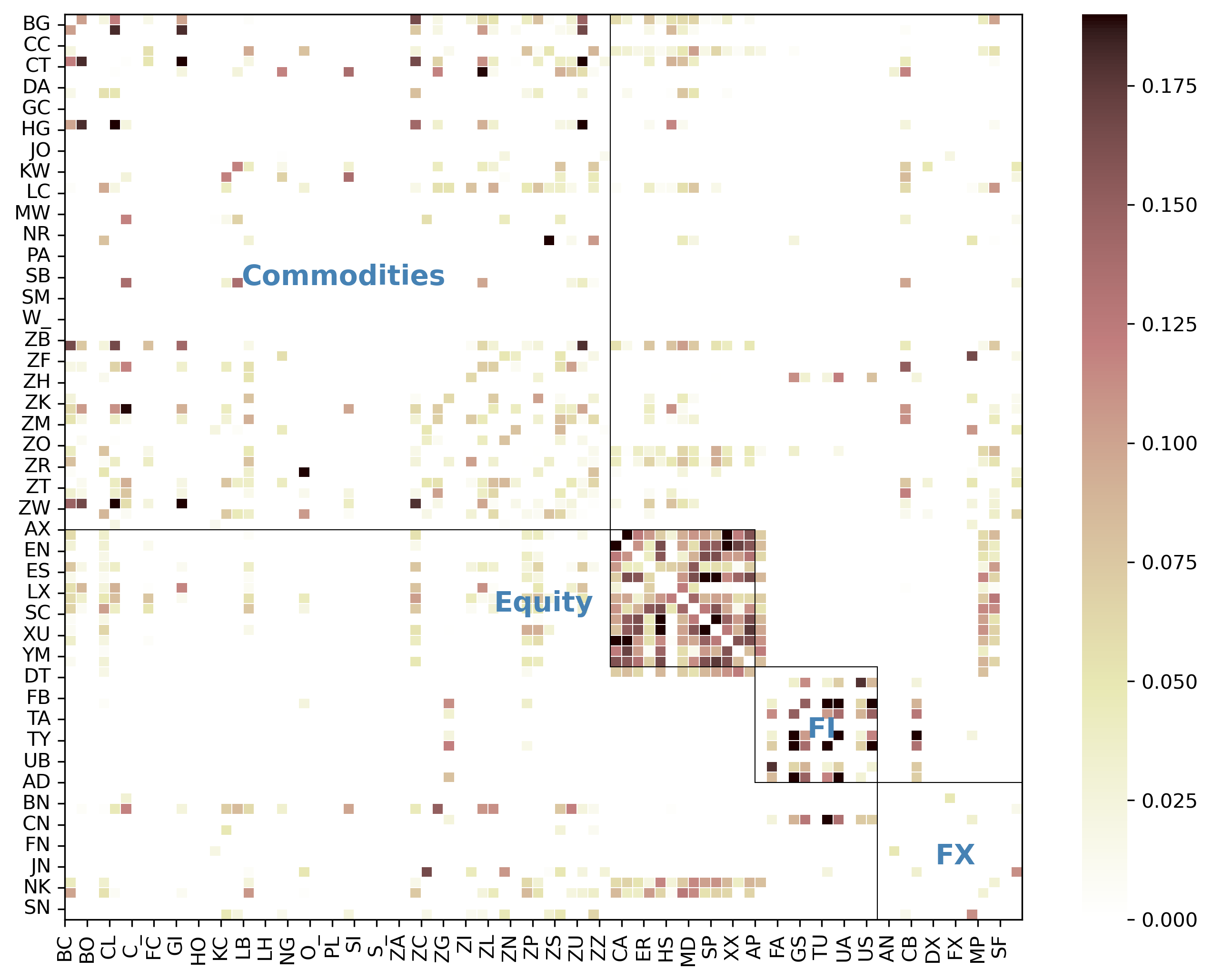}
        \caption{$\delta = 252$}
        \label{subfig:graph_gmom_lb252}
    \end{subfigure}%
    \begin{subfigure}[]{0.195\textwidth}
        \includegraphics[width=1\textwidth]{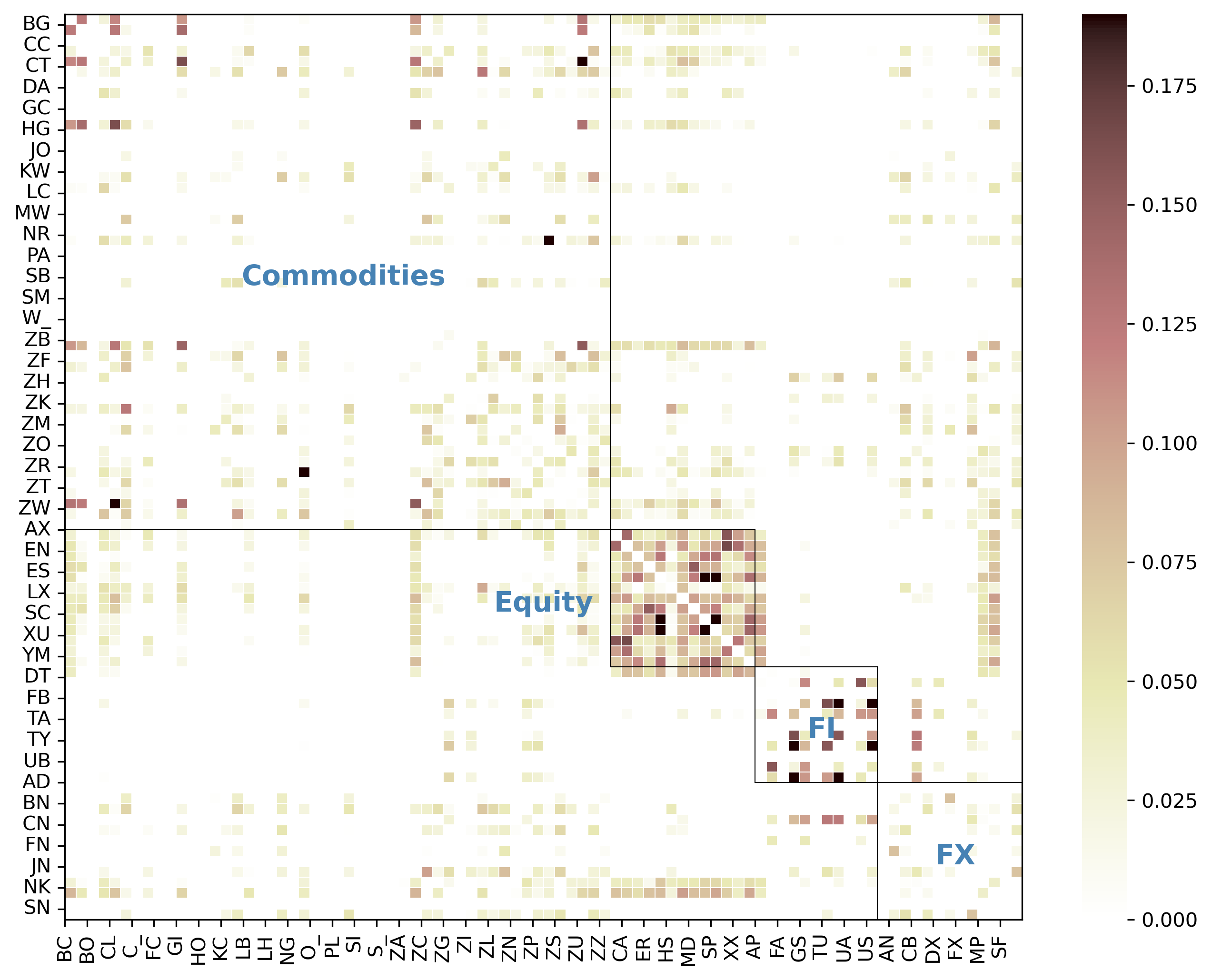}
       \caption{$\delta = 504$}
        \label{subfig:graph_gmom_lb504}
    \end{subfigure}%
    \begin{subfigure}[]{0.195\textwidth}
        \includegraphics[width=1\textwidth]{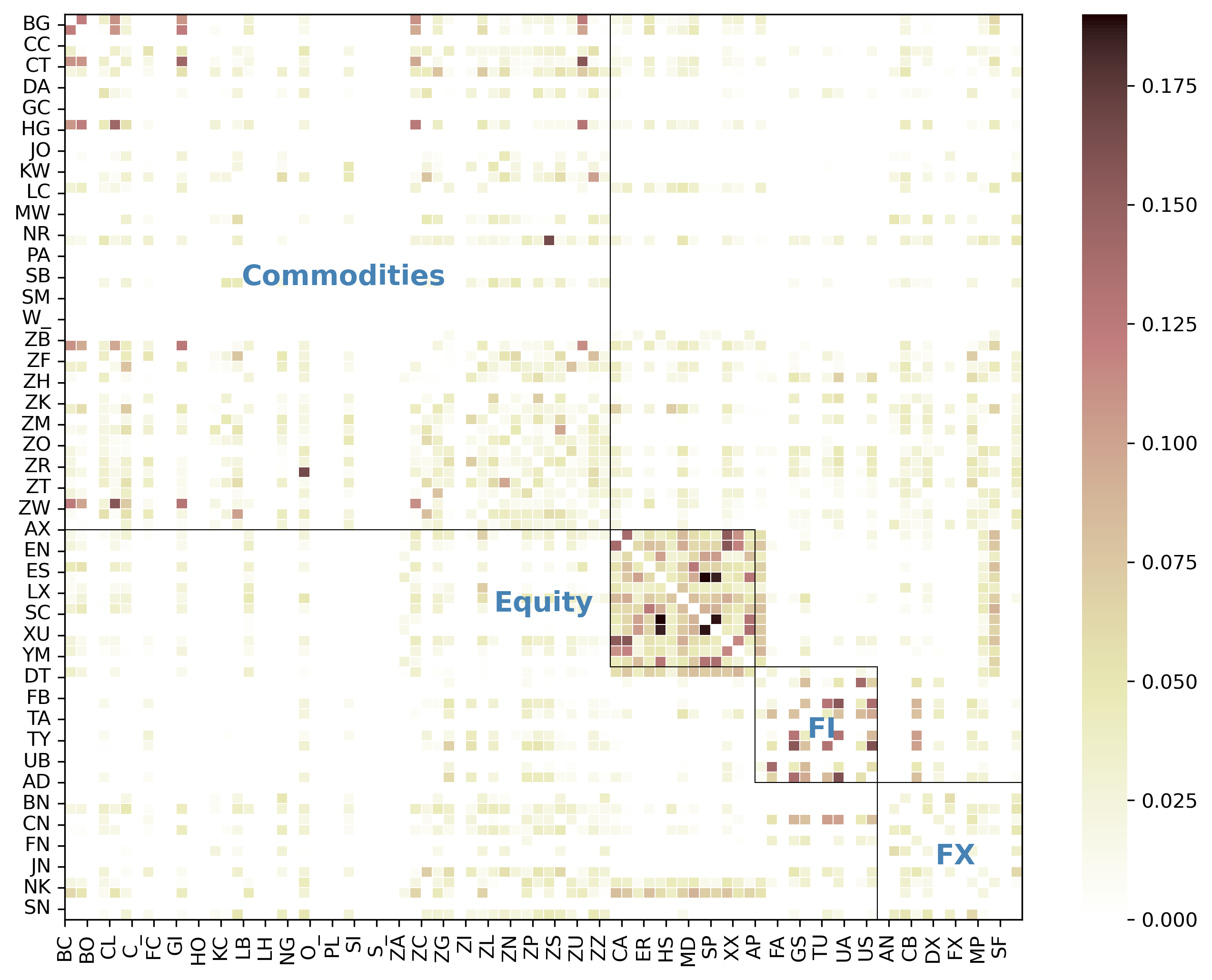}
        \caption{$\delta = 756$}
        \label{subfig:graph_gmom_lb756}
    \end{subfigure}%
    \begin{subfigure}[]{0.195\textwidth}
        \includegraphics[width=1\textwidth]{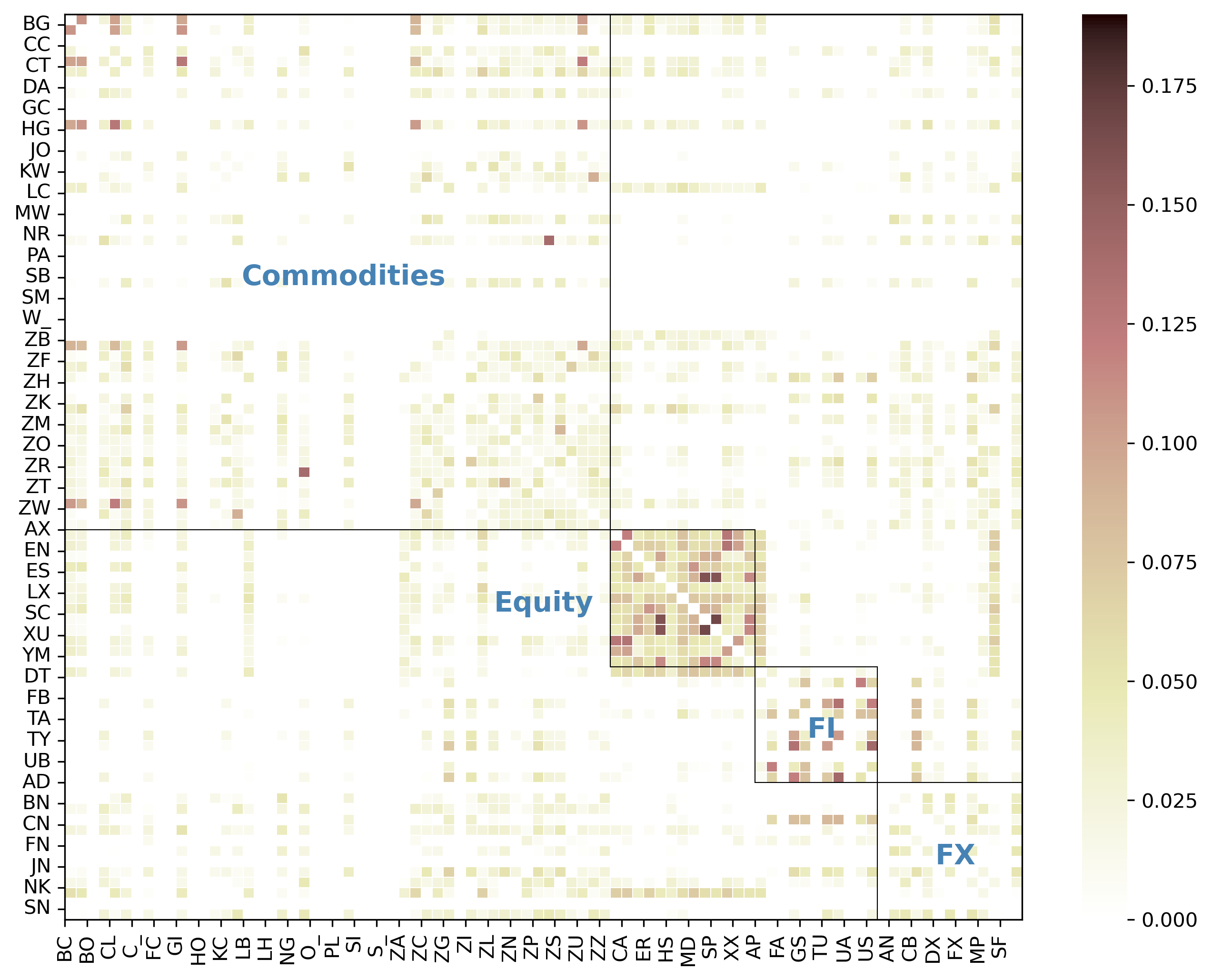}
        \caption{$\delta = 1008$}
        \label{subfig:graph_gmom_lb1008}
    \end{subfigure}%
    \begin{subfigure}[]{0.195\textwidth}
    \includegraphics[width=1\textwidth]{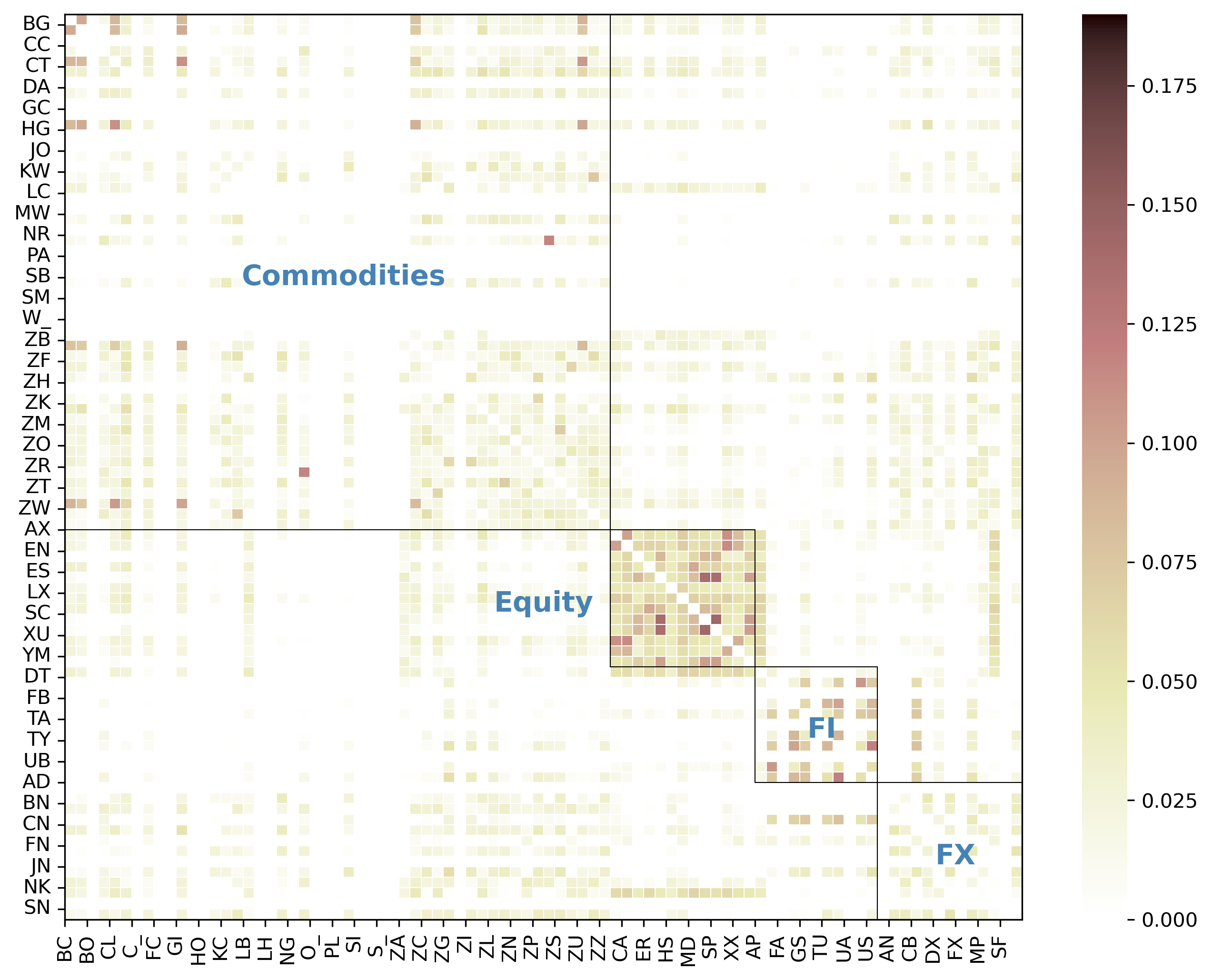}
    \caption{$\delta = 1260$}
    \label{subfig:graph_gmom_lb1260}
    \end{subfigure}%
\caption{Networks learned for 2020-03-23, obtained from different lookback windows.}
\label{fig:graph-5-lb}
\end{figure*}

\begin{figure}[h!]
\centering
    \begin{subfigure}[]{0.333\textwidth}
        \includegraphics[width=1\textwidth]{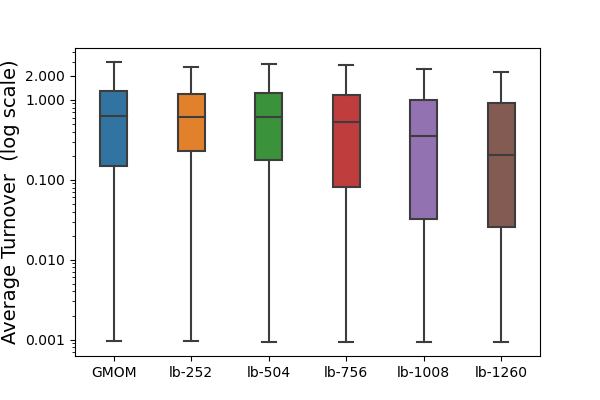}
        \caption{Average Turnover}
        \label{subfig:lb_turnover}
    \end{subfigure}%
    \begin{subfigure}[]{0.333\textwidth}
        \includegraphics[width=1\textwidth]{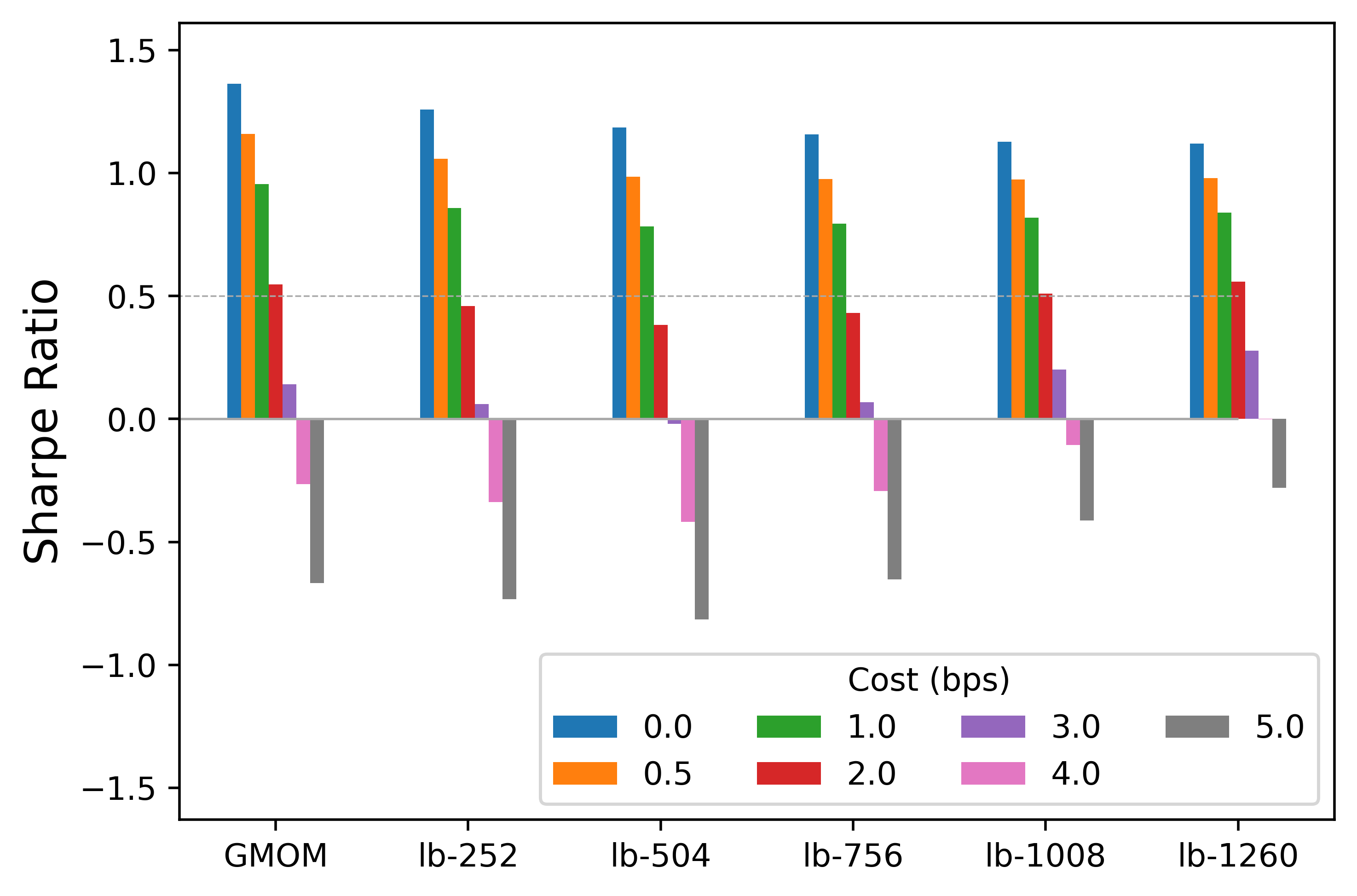}
       \caption{Cost-adjusted Sharpe Ratio}
        \label{subfig:lb_sr}
    \end{subfigure}%
    \begin{subfigure}[]{0.333\textwidth}
        \includegraphics[width=1\textwidth]{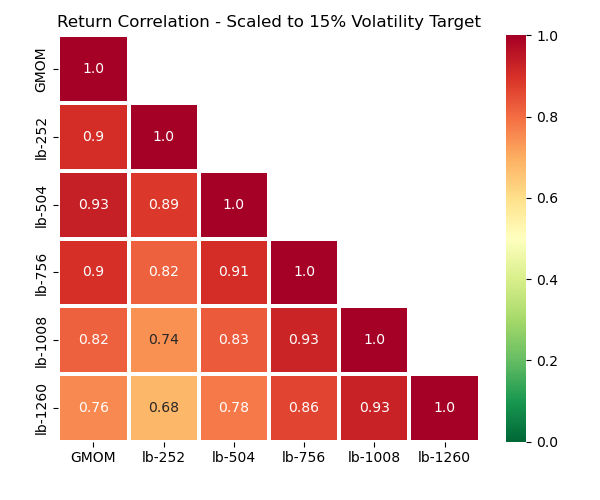}
        \caption{Scaled Return Correlation}
        \label{subfig:lb_correlation}
    \end{subfigure}
\caption{The performance of portfolios constructed from different lookback windows in graph learning}
\label{fig:lb-corr}
\end{figure}

\begin{table}[h!]
\small
\centering

\begin{threeparttable}

\caption{GMOM portfolios with different lookback windows $\delta$ - signals rescaled to 15\% target volatility}
\label{tab:bt_perf_table_scaled_lookback}
\begin{tabular}{lccccccccccc}
\toprule
{} &  return &    vol. &  Sharpe &  \makecell{downside \\ deviation} & \makecell{MDD} &  \makecell{MDD \\ duration} &  Sortino &  Calmar &  hit rate &  $\frac{\text{Avg. P}}{\text{Avg. L}}$ \\
\midrule
GMOM    & \bf{0.222} & 0.147 &   \bf{1.511} &               0.092 &         \bf{0.199} &                  \bf{6.9\%} &    \bf{2.422} &   \bf{1.179} &    \bf{55.2\%} &                 1.038 \\
$\delta$ = 252  & 0.211 & 0.147 &   1.436 &               0.094 &         0.236 &                  7.0\% &    2.241 &   0.937 &    54.7\% &                 1.048 \\
$\delta$ = 504  & 0.193 & 0.147 &   1.316 &               0.093 &         0.240 &                  7.1\% &    2.084 &   0.832 &    53.9\% &                 \bf{1.059} \\
$\delta$ = 756  & 0.189 & 0.147 &   1.288 &               0.092 &         0.278 &                  8.5\% &    2.050 &   0.702 &    54.3\% &                 1.034 \\
$\delta$ = 1008 & 0.180 & 0.147 &   1.224 &               0.093 &         0.313 &                 11.5\% &    1.938 &   0.588 &    54.1\% &                 1.031 \\
$\delta$ = 1260 & 0.182 & 0.147 &   1.235 &               0.092 &         0.287 &                 11.1\% &    1.968 &   0.649 &    54.1\% &                 1.034 \\
\bottomrule
\end{tabular}

\begin{tablenotes}
  \item[a] Best performance is in bold. No comparison of vol. and downside deviation for volatility-scaled signals. 
\end{tablenotes}
\end{threeparttable}
\end{table}

In Eq.\eqref{eq: graph_ensemble}, we employ an ensemble approach by averaging five graph adjacency matrices, each learned from momentum features with different lookback windows. This section delves into the implications of this approach and its impact on portfolio performance.  Table \ref{tab:bt_perf_table_scaled_lookback} presents a comprehensive comparison of the performance metrics for GMOM portfolios constructed with different lookback windows, with all signals rescaled to a target volatility of 15\%.  \par

The GMOM with ensemble graph exhibits superior profitability and risk tolerance capabilities, as evidenced by the highest return of 22.2\%, the highest Sharpe ratio of 1.511, and the lowest MDD of 19.9\%. The profitability appears to be primarily driven by the most recent-year graph ($\delta = 252$), while the return (scaled with respect to the target annual volatility) diminishes as we incorporate more lookback history. Concurrently, the MDD also increases. The performance of $\delta = 1260$ slightly surpasses that of $\delta = 1008$, but with very similar performance and a high correlation of 0.93, suggesting that they likely contain similar information. This indicates that graphs learned from momentum features of more than four years ago may have less informative. The correlation analysis in Figure \ref{subfig:lb_correlation} reveals that GMOM exhibits the highest correlation with the graph having a lookback window of 504, and slightly lower correlations with those of 1008 and 1260. As anticipated, the ensemble approach outperforms, emphasising the important edges captured across different windows and reducing the variance in the learned graphs. 

Interestingly, as the lookback window expands, the turnover decreases, and the decay in the cost-adjusted Sharpe ratio slows down, see Figure \ref{subfig:lb_turnover} and Figure \ref{subfig:lb_sr}. For $\delta = 1260$, the Sharpe ratio remains positive and surpasses others even after 3 bps. The Sharpe ratio exhibits a convex shape at costs of 2 and 3 bps. This could be attributed to the fact that, as shown in Figure \ref{fig:graph-5-lb}, the edge weights become more similar and the graphs denser as the lookback windows increase. Consequently, after the propagation of momentum, the network momentum might exhibit more similar values of each assets. \par

\subsection{Momentum and Reversals}
\label{sec:robust_analysis_reversal}

\begin{table}[h!]
\centering
\caption{Coefficients of Regression Models}
\label{tab:coefs}
\footnotesize
\begin{threeparttable}
\begin{tabular}{llllllllll}
\toprule
\makecell{in-sample \\ period}  & \makecell{vol-scaled \\ 1d ret} &  \makecell{vol-scaled \\ 1m return} & \makecell{vol-scaled \\ 3m return} & \makecell{vol-scaled \\ 6m return} & \makecell{vol-scaled \\ 1y return} & macd(8,24) &  macd(16,48) &     macd(32,96) \\
\midrule
 
\multicolumn{9}{c}{Panel A: LinReg (Individual Momentum Strategy)}  \\
\midrule

1990-1999 & \makecell{0.038* \\ (0.003)} & \makecell{\textcolor{Mahogany}{-0.009} \\ (0.006)} & \makecell{0.007 \\ (0.006)} & \makecell{\textcolor{Mahogany}{-0.021}* \\ (0.006)} & \makecell{0.030* \\ (0.006)} & \makecell{0.033* \\ (0.009)} & \makecell{\textcolor{Mahogany}{-0.020} \\ (0.011)} & \makecell{0.013 \\ (0.008)} \\ [0.4cm]

1990-2004 & \makecell{0.029* \\ (0.003)} & \makecell{\textcolor{Mahogany}{-0.011}* \\ (0.005)} & \makecell{0.008 \\ (0.005)} & \makecell{\textcolor{Mahogany}{-0.014}* \\ (0.005)} & \makecell{0.031* \\ (0.004)} & \makecell{0.034* \\ (0.007)} & \makecell{\textcolor{Mahogany}{-0.033}* \\ (0.008)} & \makecell{0.020* \\ (0.006)} \\ [0.4cm]

1990-2009 & \makecell{0.019* \\ (0.002)} & \makecell{\textcolor{Mahogany}{-0.005} \\ (0.004)} & \makecell{0.003 \\ (0.004)} & \makecell{\textcolor{Mahogany}{-0.006} \\ (0.004)} & \makecell{0.026* \\ (0.004)} & \makecell{0.021* \\ (0.006)} & \makecell{\textcolor{Mahogany}{-0.024}* \\ (0.007)} & \makecell{0.015* \\ (0.005)} \\ [0.4cm]

1990-2014 & \makecell{0.017* \\ (0.002)} & \makecell{\textcolor{Mahogany}{-0.004} \\ (0.003)} & \makecell{0.003 \\ (0.003)} & \makecell{\textcolor{Mahogany}{-0.002} \\ (0.003)} & \makecell{0.016* \\ (0.003)} & \makecell{0.020* \\ (0.005)} & \makecell{\textcolor{Mahogany}{-0.018}* \\ (0.006)} & \makecell{0.013* \\ (0.004)} \\ [0.4cm]

1990-2019 & \makecell{0.013* \\ (0.002)} & \makecell{\textcolor{Mahogany}{-0.002} \\ (0.003)} & \makecell{0.005 \\ (0.003)} & \makecell{0.001 \\ (0.003)} & \makecell{0.018* \\ (0.003)} & \makecell{0.013* \\ (0.005)} & \makecell{\textcolor{Mahogany}{-0.022}* \\ (0.005)} & \makecell{0.013* \\ (0.004)} \\

\midrule
\multicolumn{9}{c}{Panel B: GMOM (Network Momentum Strategy) }  \\
\midrule

1990-1999 & \makecell{0.023* \\ (0.003)} & \makecell{\textcolor{Mahogany}{-0.010} \\ (0.007)} & \makecell{0.001 \\ (0.007)} & \makecell{\textcolor{Mahogany}{-0.013} \\ (0.007)} & \makecell{0.017* \\ (0.008)} & \makecell{0.038* \\ (0.012)} & \makecell{\textcolor{Mahogany}{-0.038}* \\ (0.018)} & \makecell{0.027 \\ (0.014)} \\ [0.4cm]

1990-2004 & \makecell{0.031* \\ (0.003)} & \makecell{\textcolor{Mahogany}{-0.020}* \\ (0.005)} & \makecell{0.002 \\ (0.005)} & \makecell{\textcolor{Mahogany}{-0.005} \\ (0.006)} & \makecell{0.027* \\ (0.006)} & \makecell{0.045* \\ (0.009)} & \makecell{\textcolor{Mahogany}{-0.062}* \\ (0.012)} & \makecell{0.031* \\ (0.009)} \\ [0.4cm]

1990-2009 & \makecell{0.026* \\ (0.002)} & \makecell{\textcolor{Mahogany}{-0.016}* \\ (0.004)} & \makecell{\textcolor{Mahogany}{-0.008}* \\ (0.004)} & \makecell{0.004 \\ (0.005)} & \makecell{0.023* \\ (0.005)} & \makecell{0.028* \\ (0.008)} & \makecell{\textcolor{Mahogany}{-0.037}* \\ (0.011)} & \makecell{0.020* \\ (0.008)} \\ [0.4cm]

1990-2014 & \makecell{0.031* \\ (0.002)} & \makecell{\textcolor{Mahogany}{-0.015}* \\ (0.004)} & \makecell{\textcolor{Mahogany}{-0.011}* \\ (0.004)} & \makecell{0.004 \\ (0.004)} & \makecell{\textcolor{Mahogany}{-0.004} \\ (0.003)} & \makecell{0.035* \\ (0.007)} & \makecell{\textcolor{Mahogany}{-0.022}* \\ (0.009)} & \makecell{0.025* \\ (0.007)} \\ [0.4cm]

1990-2019 & \makecell{0.028* \\ (0.002)} & \makecell{\textcolor{Mahogany}{-0.013}* \\ (0.004)} & \makecell{\textcolor{Mahogany}{-0.004} \\ (0.003)} & \makecell{0.009* \\ (0.003)} & \makecell{\textcolor{Mahogany}{-0.001} \\ (0.003)} & \makecell{0.027* \\ (0.006)} & \makecell{\textcolor{Mahogany}{-0.033}* \\ (0.009)} & \makecell{0.029* \\ (0.006)} \\ 

\bottomrule
\end{tabular}

\begin{tablenotes}
  \item[*] Significant coefficients with $p$-value of statistic $t$ test are marked with $\ast$; the standard error is included in the parenthesis; the negative values are highlighted in red.
\end{tablenotes}

\end{threeparttable}
\end{table}

In the realm of individual momentum strategies, the phenomena of 1-month or 2-month reversals have been documented \cite{jegadeeshEvidencePredictableBehavior1990a, daCloserLookShortTerm2014, kellyUnderstandingMomentumReversal2021}. These strategies typically calculate individual momentum based on past raw returns from $t$-12 to $t$-2 months. However, our network momentum strategy is updated daily, which differs from most of previous publications that updates monthly or yearly. Moreover, there lacks a unified understanding of the persistence of such reversals within the realm of network momentum. Certain research has identified significant alpha from a signal that combines the short-term reversal with network momentum \cite{israelsenDoesCommonAnalyst2016, shahrurReturnPredictabilitySupply2010}. By contrast, studies \cite{moskowitzIndustriesExplainMomentum1999, aliSharedAnalystCoverage2020, parsonsGeographicLeadLagEffects2020} presents little to no evidence of reversals in network momentum. To account for potential reversals, we have adopted an Ordinary Least Squares (OLS) linear regression model to combine the network momentum features calculated from past returns of different periods: 1 day, 1 month, 3 months, 6 months, 1 year, and three MACD indicators of different time scales. The regression coefficients obtained from this model provide valuable insights into the relative importance of these features and also into reversal effects in network momentum.

Table \ref{tab:coefs} illustrates the coefficients and their standard errors obtained for the individual momentum strategy (Panel A) and the proposed network momentum strategy GMOM (Panel B). We adopted a rolling fashion to update our model, resulting in five regression models for the entire backtest period. Standard errors measure the standard deviation of the coefficients, enabling us to obtain $t$-statistics. Coefficients that are significant at a $p$-value of less than 0.05 are marked with an asterisk ($\ast$). Before inputting these features into the regression model, we have performed standardisation by removing the mean and scaling to unit variance. This ensures that the coefficients are comparable in scale across different features.

Several key observations can be gleaned from Table \ref{tab:coefs}. The past 1-day returns, in both individual momentum (Panel A) and network momentum (Panel B), are positive and significant, underscoring their importance in constructing daily momentum strategies. The past 1-month returns in both individual momentum (Panel A) and network momentum (Panel B) are negative, suggesting a reversal, which aligns with the findings in the literature \cite{kellyUnderstandingMomentumReversal2021}. Interestingly, past 3-month returns, while positive but non-significant in individual momentum, turn negative after the 2010 backtest period in the network momentum, a trend also observed with past 1-year returns. This suggests that network momentum might have different reversal effects than individual momentum. The MACD indicators suggest short-term and long-term momentum but indicate a mid-term reversal, which is consistent in both individual and network momentum strategies. There are advanced deep learning models designed to address changing reversals in momentum features, such as those proposed by Wood et al. \cite{woodSlowMomentumFast2022, woodTradingMomentumTransformer2022}. We acknowledge that exploring the capabilities of these models is a potential avenue for future research.

\section{Conclusion}
 

In this paper, we investigate a novel risk premium - \textit{Network Momentum} - across asset classes. We use a graph learning model to discover momentum spillover across various asset classes, focusing on the intricate interconnections within and beyond these classes to construct network momentum signals. This method significantly broadens the conventional understanding of momentum spillover, which typically focuses on pairwise relationships among companies.

Specifically, we learn dynamic networks from 64 continuous futures contracts spanning commodities, equities, fixed income, and foreign currencies. Each asset is symbolised as a node, with the adjacency matrix indicating the momentum feature similarities that shape our network. This approach reveals complex momentum spillover patterns across a wide range of assets, forming the basis of our network momentum signals. We build a long/short portfolio using these signals, demonstrating substantial profitability and a moderate correlation with individual momentum strategies. To validate the strategy's robustness, we conduct a robustness analysis highlighting the importance of inter-class connections in forming network momentum signals.

There are several paths for future research. From a modelling perspective, we could investigate whether other machine learning models might better capture the nonlinearity and temporal dynamics in network momentum, and additionally handle the situation when these features do not span an orthogonal space. We also propose investigating turnover regularisation as a potential improvement, which could enhance the monetisation of our strategy. From the perspective of financial econometrics, we recommend a comprehensive theoretical analysis of network momentum derived from pricing data. This could lead to a deeper understanding of the dynamics and potential advantages of this novel type of risk premium in financial markets.
 
\addcontentsline{toc}{section}{References}
\bibliographystyle{rQUF}
\bibliography{ref}

\clearpage
\appendix
\renewcommand\thefigure{\thesection.\arabic{figure}}    
\renewcommand\thetable{\thesection.\arabic{table}}

\section{Appendix}

\subsection{Dataset Details}
\label{sec:appendix-dataset} 
In Table \ref{table:universe}, we list the ticker name of all the future contracts in our universe. 
  
\begin{table}[h!]
\centering
\footnotesize
\caption{The Pinnacle Universe}
\label{table:universe}
\begin{tabular}{lll|lll}
\toprule
\textbf{Ticker} & \textbf{Description} & \textbf{Period} & \textbf{Ticker} & \textbf{Description} & \textbf{Period}\\
\midrule
\multicolumn{2}{l}{\textbf{Commodities:}  } & & \multicolumn{2}{l}{\textbf{Equities:}  } \\
BC & BRENT CRUDE OIL, composite & 2010-2022 & AX & GERMAN DAX INDEX & 1999-2022 \\
BG & BRENT GASOIL, composite & 2010-2022 & CA & CAC40 INDEX & 2000-2022 \\
CC & COCOA & 1990-2022 & EN & NASDAQ, MINI & 2001-2022 \\
CL & CRUDE OIL & 1990-2021 & ER & RUSSELL 2000, MINI & 2004-2022 \\
CT & COTTON \#2 & 1990-2022 & ES & S\&P 500, MINI & 1999-2022 \\
DA & MILK III, composite & 1999-2022 & HS & HANG SENG INDEX & 1999-2022 \\
GI & GOLDMAN SAKS C. I. & 1995-2022 & LX & FTSE 100 INDEX & 1991-2022\\
JO & ORANGE JUICE & 1990-2022 & MD & S\&P 400, MINI, electronic & 1994-2022 \\
KC & COFFEE & 1990-2022 & SC & S\&P 500, composite & 1996-2022 \\
KW & WHEAT & 1990-2022 & SP & S\&P 500, day session & 1990-2022 \\
LB & LUMBER & 1990-2022 & XU & DOW JONES EUROSTOXX 50 & 2003-2022 \\
MW & WHEAT, MINN & 1990-2022 & XX & DOW JONES STOXX 50 & 2004-2022 \\
NR & NATURAL GAS & 1990-2021 & YM & DOW JONES, MINI (\$5.00) & 2004-2022 \\
SB & SUGAR \#11 & 1990-2022  & \multicolumn{2}{l}{\textbf{Fixed Income:}  }  \\
W\_ & WHEAT, CBOT & 1990-2018 & AP &  AUSTRALIAN PRICE INDEX & 2010-2022 \\
ZA & PALLADIUM, electronic &1990-2022  & DT &  EURO BOND (BUND) & 1991-2022\\
ZB & RBOB, electronic & 1990-2022 & FB &  T-NOTE, 5yr composite &1990-2022\\
ZC & CORN, electronic & 1990-2022 & GS &  GILT, LONG BOND & 1991-2022\\
ZF & FEEDER CATTLE, electronic & 1990-2022 & TU &  T-NOTES, 2yr composite & 1992-2022\\
ZG & GOLD, electronic & 1990-2022 & TY &  T-NOTE, 10yr composite & 1990-2022\\
ZI & SILVER, electronic &1990-2022 & UB &  EURO BOBL & 2000-2022\\
ZK & COPPER, electronic & 1990-2022 & US &  T-BONDS, composite & 1990-2022 \\
ZL & SOYBEAN OIL, electronic &1990-2022 & \multicolumn{2}{l}{\textbf{Currencies:}  }  \\
ZM & SOYBEAN MEAL, electronic &1990-2022 & AN & AUSTRALIAN \$\$, day session & 1990-2022 \\
ZN & NATURAL GAS, electronic & 1992-2022 & BN & BRITISH POUND, composite & 1990-2022\\
ZO & OATS, electronic &1990-2022  & CB & CANADIAN 10YR BOND & 1996-2022\\
ZP & PLATINUM, electronic & 1990-2022 & CN & CANADIAN \$\$, composite & 1990-2022\\
ZR & ROUGH RICE, electronic & 1990-2022 & DX & US DOLLAR INDEX & 1990-2022\\
ZS & SOYBEANS electronic & 1990-2022 & FN & EURO, composite &1990-2022\\
ZT & LIVE CATTLE, electronic & 1990-2022 & JN & JAPANESE YEN, composite & 1990-2022\\
ZU & CRUDE OIL, electronic & 1990-2022 & MP & MEXICAN PESO & 1997-2022\\
ZW & WHEAT electronic & 1990-2022 & NK & NIKKEI INDEX & 1992-2022\\
ZZ & LEAN HOGS, electronic & 1990-2022 & SN & SWISS FRANC, composite &1990-2022 \\

\bottomrule
\end{tabular}
\end{table}

\subsection{Supplementary Definitions}
\label{sec:appendix-math-preprocs}

\paragraph{Exponential Weighted Moving standard deviation (EWMstd)} 
Given a series of daily returns $r_t$, the exponential weighted moving standard deviation $\sigma_t$ with a span of $N$ days can be defined as:
\begin{align*}
\alpha & = \frac{2}{{N + 1}} \\
w_t & = (1 - \alpha)^t \\
\mu_t & = \frac{{\sum_{\tau=0}^{t} w_{\tau} \cdot r_{t-\tau}}}{{\sum_{\tau=0}^{t} w_{\tau}}} \\
\sigma_t & = \sqrt{\frac{{\sum_{\tau=0}^{t} w_{\tau} \cdot (r_{t-\tau} - \mu_t)^2}}{{\sum_{\tau=0}^{t} w_{\tau}}}}
\end{align*}
Where:
\begin{itemize}
\setlength\itemsep{-0.5em}
\item \( \alpha \) is the smoothing factor;
\item \( w_t \) are the weights, decaying exponentially;
\item \( \mu_t \) is the exponential weighted moving average (EWMA) of the returns;
\item \( \sigma_t \) is the exponential weighted moving standard deviation of the returns;
\item \( N \) is the span, which is 60 in your specific case.
\end{itemize}

\paragraph{Exponentially Weighted Moving Average $m(i, t, J)$ in Eq.\eqref{eq:macd_ewm}}

Let \( p(i,t) \) denote the price of asset \(i\) at time \(t\), and let \( J \) be the scale parameter for the exponentially weighted moving average. Then, the exponentially weighted moving average \( m(i,t,J) \) is given by:
\begin{align*}
\alpha & = \frac{1}{J} \\
m(i,t,J) & = \alpha \cdot p(i,t) + (1 - \alpha) \cdot m(i,t-1,J)
\end{align*}
where the initial value of \( m \) can be chosen as the first price or any other method of initialisation, $\alpha$ is the smoothing factor, and the half-life decay rate is defined as:
\[
HL = \frac{{\log(0.5)}}{{\log(1-1/J)}}
\]
This formulation provides a weight decay that is controlled by the scale \( J \), such that the influence of past prices declines exponentially with time.
\end{document}